\documentclass[twocolumn]{aastex631}

\usepackage{graphicx}	
\usepackage{amsmath}	
\usepackage{amssymb}	
\usepackage{comment}   
\usepackage{booktabs}  %
\usepackage{hyperref}
\usepackage{ragged2e}
\usepackage{float}
\usepackage{multirow}

\begin{document}

\title{Multi-wavelength observations of multiple eruptions of the recurrent nova M31N~2008-12a}
\shorttitle{Multiple eruptions of RN M31N 2008-12a}
\shortauthors{Basu, J. et al.}
 
\correspondingauthor{Judhajeet Basu}
\email{judhajeet.basu@iiap.res.in}

\author[0000-0001-7570-545X]{Judhajeet Basu}
\affiliation{Indian Institute of Astrophysics, 2nd Block Koramangala, 560034, Bangalore, India}
\affiliation{Pondicherry University, R.V. Nagar, Kalapet, 605014, Puducherry, India}

\author[0000-0002-0000-1543]{M. Pavana}
\affiliation{Indian Institute of Astrophysics, 2nd Block Koramangala, 560034, Bangalore, India}
\affiliation{Pondicherry University, R.V. Nagar, Kalapet, 605014, Puducherry, India}

\author[0000-0003-3533-7183]{G.C. Anupama}
\affiliation{Indian Institute of Astrophysics, 2nd Block Koramangala, 560034, Bangalore, India}

\author[0000-0002-3927-5402]{Sudhanshu Barway}
\affiliation{Indian Institute of Astrophysics, 2nd Block Koramangala, 560034, Bangalore, India}

\author[0000-0001-6952-3887]{Kulinder Pal Singh}
\affiliation{Department of Physical Sciences, Indian Institute of Science Education and Research Mohali, 140306, Punjab, India}

\author[0000-0002-7942-8477]{Vishwajeet Swain}
\affiliation{Physics Department, Indian Institute of Technology Bombay, Powai, Mumbai, 400076, India}

\author[0000-0003-4524-6883]{Shubham Srivastav}
\affiliation{Astrophysics, Department of Physics, University of Oxford, Keble Road, Oxford, OX1 3RH, UK}
\affiliation{Astrophysics Research Centre, School of Mathematics and Physics, Queen's University Belfast, Belfast BT7 1NN, UK}

\author[0000-0003-0871-4641]{Harsh Kumar}
\affiliation{Harvard College Observatory, Harvard University, 60 Garden St. Cambridge 02158 MA, USA}
\affiliation{Center for Astrophysics, Harvard University, 60 Garden St. Cambridge 02158 MA, USA }
\affiliation{Physics Department, Indian Institute of Technology Bombay, Powai, Mumbai, 400076, India}

\author[0000-0002-6112-7609]{Varun Bhalerao}
\affiliation{Physics Department, Indian Institute of Technology Bombay, Powai, Mumbai, 400076, India}

\author[0000-0002-2033-3051]{L. S. Sonith}  
\affiliation{Indian Institute of Astrophysics, 2nd Block Koramangala, 560034, Bangalore, India}
\affiliation{Pondicherry University, R.V. Nagar, Kalapet, 605014, Puducherry, India}

\author{G. Selvakumar}
\affiliation{Indian Institute of Astrophysics, 2nd Block Koramangala, 560034, Bangalore, India}

\begin{abstract}
We report the optical, UV, and soft X-ray observations of the $2017-2022$ eruptions of the recurrent nova M31N~2008-12a. We find a ``cusp'' feature in the $r'$ and $i'$ band light curves close to the peak, which could be related to jets. The geometry of the nova ejecta based on morpho-kinematic modelling of the H$\alpha$ emission line indicates an extended jet-like bipolar structure. Spectral modelling indicates an ejecta mass of 10$^{-7}$ to 10$^{-8}$ M$_{\odot}$ during each eruption and an enhanced Helium abundance. The super-soft source (SSS) phase shows significant variability, which is anti-correlated to the UV emission, indicating a common origin. The variability could be due to the reformation of the accretion disk. We infer a steady decrease in the accretion rate over the years based on the inter-eruption recurrence period. A comparison of the accretion rate with different models on the $\rm M_{WD}$$-\dot{M}$ plane yields the mass of a CO WD, powering the ``H-shell flashes'' every $\sim$~1~year, to be $>1.36$ M$_{\odot}$ and growing with time, making M31N~2008-12a a strong candidate for the single degenerate scenario of Type Ia supernovae progenitor.

\end{abstract}


\section{Introduction}
\label{intro}

Nova eruptions are a consequence of thermonuclear runaway on the surface of a white dwarf (WD) primary in cataclysmic binary systems, resulting in the ejection of material in the range of $\rm~10^{-7}~-~10^{-4}~M_\odot$ (\citealt{Geh98,Her98} and \citealt{Sta99}). Inherently, all novae are supposed to be recurrent, with the primary WD and the secondary red-giant/sub-giant star sustaining all the eruptions. The observed recurrence period of novae can range from 1~year (M31N~2008-12a; \citealt{Dar14}) to 98~years (V2487~Ophiuchi; \citealt{Sch10}).   

M31N~2008-12a is an extraordinary RN whose eruptions have been observed every year in 2008--2023 (Table~\ref{tab:eruption_all}). It was first discovered during its 2008 eruption by \cite{nis08}, though previous eruptions in 1992, 1993 and 2001 have been retrieved from the archives. Since the 2013 eruption, it has been monitored and studied across different wavelength ranges to understand its short recurrence period (2013 eruption -- \citealt{Dar14,Hen14,tan14}; 2014 -- \citealt{dar15,hen15}; 2015 -- \citealt{Dar16a,dar17a,dar17b}; 2016 -- \citealt{Hen18}). 

The optical light curve and spectral evolution of this very fast RN were found to be similar during all the eruptions, with Balmer, He, and N lines dominating the spectrum. Light curves showed an extremely rapid rise to maximum ($\sim 1$~day) followed by a fast linear decline for about 4~days and a plateau with slow decline and jitters from day 4 to 8. The multi-eruption UV light curves were similar, with an initial rapid linear decline followed by slow plateau-like declines. The plateau phase coincided with the supersoft X-ray source (SSS) phase. Like optical and UV light curves, the SSS phase was similar during multiple eruptions. 

The 2016 eruption \citep{Hen18}, however, deviated from the general trend. It occurred after a longer inter-eruption gap, the optical light curve showed a short-lived cuspy feature, and the UV and X-ray fluxes disappeared relatively early compared to the previous eruptions. The ``peculiar" behaviour of the 2016 eruption was suggested to be due to a lower accretion rate prior to the 2016 eruption. 

Theoretical models generated to satisfy the short recurrence period and short turn-on time of the SSS phase have allowed constraining the mass of the WD to near the Chandrasekhar limit \citep{tan14,kato14}. 

Deep H$\alpha$ and HST imaging revealed the presence of an elliptical ($134 \times 90$~parsecs) super-remnant nebula around M31N~2008-12a \citep{dar15}. The size and mass of the shell indicate that the system has been undergoing eruptions for $\sim 10^6$ years \citep{dar19} and would likely do so for $\sim 2\times 10^4$ more years before the WD attains $\rm M_{Ch}$ \citep{dar17b}.

This paper discusses the optical photometric and spectroscopic observations during the $2017-2022$ eruptions and the evolution of UV and soft X-ray emission based on \textit{Swift} archival data and data from our observations with the \textit{AstroSat}. We explore spectroscopic modelling to reveal physical processes during outbursts and derive associated physical parameters. The behaviour of the RN M31N~2008-12a during the $2017-2022$ eruptions is also compared to that of the previous eruptions. We end with a discussion on the recurrence period and its implication on the accretion rate and the mass of the primary WD.

\begin{table*}
\centering
\caption{All known eruption dates of M31N 2008-12a till 2022}
\hspace*{-2.cm}
\begin{tabular}{llllll}
\toprule
Eruption date$^{(a)}$ & Discovery & SSS-on date$^{(b)}$ & Days since                & Detection wavelength & References \\
(UT)                  & mag (Filter)      & (UT)                & last eruption & (Observatory) &          \\
\midrule
(1992 Jan 28)            &  $\cdots$    & 1992 Feb 03              & $\cdots$        & X-ray (ROSAT)                       & 1, 2\\
(1993 Jan 03)            &  $\cdots$    & 1993 Jan 09              & 341             & X-ray (ROSAT)                       & 1, 2\\
(2001 Aug 27)            &  $\cdots$    & 2001 Sep 02              & $\cdots$        & X-ray (Chandra)                     & 2, 3 \\
2008 Dec 25              &  $\cdots$    & $\cdots$                 & $\cdots$        & Visible (Miyaki-Argenteus)          & 4 \\
2009 Dec 02              &  $\cdots$    & $\cdots$                 & 342             & Visible (PTF)                       & 5 \\
2010 Nov 19              &  $\cdots$    & $\cdots$                 & 352             & Visible (Miyaki-Argenteus)          & 2 \\
2011 Oct 22.5            &  $\cdots$    & $\cdots$                 & 337.5           & Visible (ISON-NM)                   & 5-8 \\
2012 Oct 18.7            &  $\cdots$    & $<$ 2012 Nov 06.45         & 362.2           & Visible (Miyaki-Argenteus)          & 8-11\\
2013 Nov 26.95 $\pm$ 0.25 &  18.9 (R)           & $\le$ 2013 Dec 03.03    & 403.5           & Visible (iPTF); UV/X-ray (Swift)    & 5, 8, 11-14\\
2014 Oct 02.69 $\pm$ 0.21 &  18.86 ($r^\prime$)         & 2014 Oct 08.6 $\pm$ 0.5 & 309.8 $\pm$ 0.7 & Visible (LT); UV/X-ray (Swift)      & 8, 15 \\
2015 Aug 28.28 $\pm$ 0.12 &  19.09 ($r^\prime$)         & 2015 Sep 02.9 $\pm$ 0.7 & 329.6 $\pm$ 0.3 & Visible (LCO); UV/X-ray (Swift)     & 14, 16-18 \\
2016 Dec 12.32 $\pm$ 0.17 &  17.62 (V)          & 2016 Dec 17.2 $\pm$ 1.1 & 471.7 $\pm$ 0.2 & Visible (Itagaki); UV/X-ray (Swift) & 19-23 \\
2017 Dec 31.58 $\pm$ 0.20 $^{\dagger}$ & 18.41 (clear) & 2018 Jan 05.6 $\pm$ 0.5 & 384.3 $\pm$ 0.4 & Visible (WCO); UV/X-ray (Swift)     & 24-27 \\
2018 Nov 06.67 $\pm$ 0.13 $^{\dagger}$ & 19.15         & 2018 Nov 13.2 $\pm$ 0.5 & 310.1 $\pm$ 0.3 & Visible (LT); UV/X-ray (Swift)      & 28-31 \\
2019 Nov 06.60 $\pm$ 0.11 $^{\dagger}$ & 19.40         & 2019 Nov 12.4 $\pm$ 0.5 & 364.9 $\pm$ 0.2 & Visible (HO); UV/X-ray (Swift)      & 32-34 \\
2020 Oct 30.49 $\pm$ 0.34 $^{\dagger}$ & 18.74 ($g^\prime$)    & 2020 Nov 05.5 $\pm$ 0.9 & 358.9 $\pm$ 0.4 & Visible (LT);  UV/X-ray (Swift)     & 35-38 \\
2021 Nov 14.17 $\pm$ 0.21 $^{\dagger}$ & 18.7 (clear) & 2021 Nov 19.2 $\pm$ 0.6 & 379.7 $\pm$ 0.5 & Visible (Itagaki); UV/X-ray (Swift) & 39-42 \\
2022 Dec 02.50 $\pm$ 0.34 $^{\dagger}$ & 19.18 ($r^\prime$)   & 2022 Dec 07.5 $\pm$ 0.5 & 383.3 $\pm$ 0.5 & Visible (LCOGT); UV/X-ray (Swift)   & 43-45 \\
2023 Dec 05.28 $\pm$ 0.07 $^{\dagger \ddag}$ & 18.63 (CV)  &    $\cdots$         & 367.8 $\pm$ 0.4 & Visible (HMT, XO) & 46-50 \\
\bottomrule
\end{tabular}
    
\label{tab:eruption_all}

\justifying
\textbf{Notes:} Updated version of Table 1 of \citealt{tan14,dar15,hen15a,Dar16a,Hen18}. \\
$^{\dagger}$ Determined in this paper. \\
$^{\ddag}$ The 2023 eruption was discovered during the revision of the manuscript. As the observations are still ongoing, all details are not yet available. \\
$^{(a)}$ Archival X-ray detections (cf., \citealt{hen15a}) are enclosed in brackets. \\
$^{(b)}$ ROSAT data was used to estimate the SSS $\rm t_{on}$ for 1992 and 1993. The Chandra detection in 2001 Sep 08 UT was taken as the midpoint of a typical 12-day SSS phase to constrain the eruption date. \\
$\cdots$ indicates unavailability of information. \\
References - (1) \cite{whi95}, (2) \cite{hen15a}, (3) \cite{wil04}, (4) \cite{nis08}, (5) \cite{tan14}, (6) \cite{kor11}, (7) \cite{bar11}, (8) \cite{dar15}, (9) \cite{nis12}, (10) \cite{sha12}, (11) \cite{Hen14}, (12) \cite{tan13}, (13) \cite{Dar14}, (14) \cite{Dar16a}, (15) \cite{hen15}, (16) \cite{dar15a}, (17) \cite{dar15b}, (18) \cite{hen15b}, (19) \cite{Hen18}, (20) \cite{ita16}, (21) \cite{ita16a}, (22) \cite{hen16}, (23) \cite{hen16a}, (24) \cite{boyd17}, (25) \cite{hen18b}, (26) \cite{hen18a}, (27) \cite{nai18},  (28) \cite{hen18c}, (29) \cite{Dar18a}, (30) \cite{Tan18}, (31) \cite{hen18d}, (32) \cite{dar19a}, (33) \cite{Oks19}, (34) \cite{dar19b}, (35) \cite{gal20}, (36) \cite{dar20}, (37) \cite{dar20a}, (38) \cite{dar20b}, (39) \cite{ita21}, (40) \cite{tan21}, (41) \cite{dar21}, (42) \cite{dar21a}, (43) \cite{per22}, (44) \cite{Sha22}, (45) \cite{dar22}, (46) \cite{sun23}, (47) \cite{shafter23}, (48) \cite{perez23}, (49) \cite{basu23}, (50) \cite{balcon23}.

\end{table*}


\section{Observations}
\subsection{Optical}

\begin{table}
    \caption{Optical spectroscopic and photometric observations of 2018--2022 eruptions of M31N~2008-12a. This is an example table. The full table will be made available in the online format}
    \label{phot_mag}
    \resizebox{\columnwidth}{!}{%
    \begin{tabular}{ccccc}
    \toprule
    Date (UT)  & Telescope & Instrument & Grating & Exp (s) \\
    \midrule
   2018 Nov 07.8 &  HCT & HFOSC & Gr7 &  2700 \\
   2018 Nov 08.6 &  HCT & HFOSC & Gr7 &  3600 \\
   2019 Nov 07.5 &  HCT & HFOSC & Gr7 &  3000 \\
   2020 Oct 31.6 &  HCT & HFOSC & Gr7 &  3600 \\
   2021 Nov 14.8 &  HCT & HFOSC & Gr7 &  2100 \\
   2021 Nov 15.6 &  HCT & HFOSC & Gr7 &  3600 \\
   2022 Dec 03.7 &  HCT & HFOSC & Gr7 &  3600 \\
   2022 Dec 04.5 &  HCT & HFOSC & Gr7 &  3600 \\
   \midrule
    Date (UT)  & Telescope & Filter & Magnitude & Exp (s)\\
    \midrule
   2018 Nov 07.58 & JCBT & $B$ & 19.29 $\pm$ 0.28 & 1800 \\
   2018 Nov 07.57 & JCBT & $V$ & 18.78 $\pm$ 0.10 & 1200 \\
   2018 Nov 08.57 & HCT & $V$ & 19.27 $\pm$ 0.13 & 300$\times$3 \\
   2018 Nov 08.56 & HCT & $R$ & 19.11 $\pm$ 0.11 & 180$\times$3 \\
   2018 Nov 08.55 & HCT & $I$ & 18.83 $\pm$ 0.20 & 150$\times$3 \\
   2018 Nov 08.76 & GIT & $g'$ & 19.60 $\pm$ 0.10 & 300$\times$3 \\
   2018 Nov 08.77 & GIT & $r'$ & 19.15 $\pm$ 0.09 & 300$\times$3 \\
   2018 Nov 08.74 & GIT & $i'$ & 19.28 $\pm$ 0.15 & 300$\times$3 \\
    
\bottomrule
\end{tabular} }
\end{table}

Photometric and spectroscopic observations were carried out using the following telescopes and instruments. The log of optical observations is given in Table~\ref{phot_mag}.

\subsubsection{GROWTH-India Telescope}
The GROWTH-India Telescope (GIT; \citealt{harsh22b} )\footnote{\href{https://sites.google.com/view/growthindia/about}{https://sites.google.com/view/growthindia/about}} is a 0.7~m fully robotic telescope at the Indian Astronomical Observatory (IAO), Hanle, India. The telescope has a 4096 $\times$ 4108 Andor iKon-XL CCD. The detector has an image scale of 0.67$''$ pixel$^{-1}$ with a field of view (FoV) of 0.7$^{\circ}$.

The GIT images were pre-processed, i.e., bias subtracted, flat-fielded and cosmic-ray corrected by the automated pipeline of the GIT \citep{harsh22b}. Multiple exposures were obtained every night, and in the case of low SNR, images from the same night and the same filter were stacked using \texttt{SWarp} \citep{Ber02_terapix}. 

For the 2018 data, aperture photometry was performed using an aperture of $\sim2.5 ''$, close to the full-width half maximum (FWHM) of the stellar profile in the images due to low SNR. For 2020--2022 data, the FWHM was first calculated using \texttt{SExtractor} \citep{Ber96_sextractor} and subsequently used in Image Reduction and Analysis Facility (\texttt{IRAF} \footnote{IRAF is distributed by the National Optical Astronomy Observatories, which are operated by the Association of Universities for Research in Astronomy, Inc., under cooperative agreement with the National Science Foundation.}, \citet{tod93}) to perform PSF photometry. The magnitudes of the local standard stars given in \cite{Dar16a} were converted from $BVRI$ to $gri$ using the transformations in \cite{Jes05} to determine the zero points for photometric calibrations. Aperture photometry was also performed and found to be consistent with PSF photometry, eliminating any systematic differences that could arise from aperture photometry of the 2018 data.

\subsubsection{Himalayan Chandra Telescope}
The Himalayan Faint Object Spectrograph Camera (HFOSC)\footnote{\href{https://www.iiap.res.in/?q=iao_2m_hfosc}{https://www.iiap.res.in/?q=iao\_2m\_hfosc}} mounted on the 2~m Himalayan Chandra Telescope (HCT) located at IAO, Hanle, India was used to obtain images in the $VRI$ bands on 2018~Nov~08~UT. HFOSC is equipped with a 2K~$\times$~4K CCD. The pixels correspond to an image scale of 0.296$''$ pixel$^{-1}$, with a FoV of $10^\prime~\times~10^\prime$ for the central 2K~$\times$~2K region. The images were pre-processed using the standard routines in \texttt{IRAF}. The instrumental magnitudes were obtained using aperture photometry with an aperture set at a radius three times the FWHM. Differential photometry was performed with respect to the local standards \citep{Dar16a} to account for the zero points of the images. 

Optical spectra obtained using HFOSC \citep{pav18, son21, basu22} are given in Table~\ref{phot_mag}. We used a grism with R~$\approx$~1200 in the wavelength range of 3500~-~7800~\AA. Data reduction was performed in the standard manner using \texttt{IRAF}. All the spectra were bias subtracted and cosmic rays were corrected before extraction. Wavelength calibration was carried out using the FeAr arc lamp spectrum. Spectro-photometric standard stars, Feige 110 (2018, 2019 and 2020 eruptions) and Feige 34 (2021 and 2022 eruptions), were used to correct for the instrumental response and bring the spectra to a relative flux scale. Absolute flux calibration was done using zero points obtained from broadband magnitudes based on photometric observations within 3-4 hours of the spectroscopic observation, except in the case of the 2020 observations, which had a gap of around 7 hours.

\subsubsection{J.C. Bhattacharyya Telescope}
The 2K $\times$ 4K UK Astronomy Technology Centre (UKATC) CCD mounted on the 1.3 m Jagadish Chandra Bhattacharya Telescope (JCBT)\footnote{\href{https://www.iiap.res.in/?q=centers/vbo\#Telescopes_VBO}{https://www.iiap.res.in/?q=centers/vbo\#Telescopes\_VBO}}, located at the Vainu Bappu Observatory (VBO), Kavalur, India, was used during the 2018 eruption. It has a 15-micron pixel size corresponding to an image scale of 0.3$''$ pixel$^{-1}$, with a FoV of $10'\times  20'$. JCBT observed the nova in $BV$ bands 2~days after the eruption. The images were reduced and calibrated following the same steps used for HCT data.

\subsubsection{Other data sources}

Our observations were combined with publicly available photometric data for analysis, from sources referenced below.  
\begin{itemize}
    \item 2017: \citet{socia18,horn18,kaur18,kaur18a,naito18,kaur18b,erdman18,darn18}.
    \item 2018: \citet{Eng18,Agni18a,Wys18,Tan18,Agni18b,Kau18}; Zwicky Transient Facility (ZTF) archive (\texttt{ALeRCE Explorer} \footnote{\href{https://alerce.online/}{https://alerce.online/}}, \citealt{alerce21}).
    \item 2019: \citet{hor19,hors19,kaur19}; ZTF archive.
    \item 2020: \citet{perez20,gal20,rajagopal20}.
    \item 2021: \citet{taguchi21, naito21}
    \item 2022: \citet{rodriguez23,Sha22,agnihotri22,tan22,erdman22}.
\end{itemize}

\subsection{Ultraviolet}

\begin{table}
    \caption{\textit{AstroSat} (UVIT and SXT) and \textit{Swift} (UVOT and XRT) observations of $2017-2022$ eruptions of the M31N 2008-12a. The complete table will be made available in online format.}
    \label{swift_tables}
    \label{astrosat_table}
    \resizebox{\columnwidth}{!}{%
    \hspace*{-1cm}
    \begin{tabular}{cccccc}
    \toprule
    Obs ID & Date & UV mag (AB) & Exp & Count rate  ($\times 10^{-3}$ count s$^{-1}$) & Exp \\
    \cmidrule(lr){1-1} \cmidrule(lr){2-2} \cmidrule(lr){3-4} \cmidrule(lr){5-6}
    \textit{AstroSat} & (UT) & UVIT F148W & (s) & SXT $(0.3-2.0)$ keV & (s) \\
    \cmidrule(lr){1-1} \cmidrule(lr){2-2} \cmidrule(lr){3-4} \cmidrule(lr){5-6}
    T03\_156T01\_9000003312 & 2019-11-19.78 & 23.03 $\pm$ 0.14 & 8548 & -- & -- \\
    T03\_259T01\_9000003972 & 2020-11-03.25 & 21.30 $\pm$ 0.09 & 4888 & -- & -- \\
    T03\_262T01\_9000003988 & 2020-11-10.76 & \multirow{7}{*}{22.78 $\pm$ 0.12} & \multirow{7}{*}{8170} & 25.81 $\pm$ 2.18 & 8909 \\
                         & 2020-11-11.09 &                   & & 34.33$\pm$ 2.68 & 7254 \\
                         & 2020-11-11.42 &                   & & 31.30 $\pm$ 2.18 & 11355 \\
                         & 2020-11-11.75 &                   & & 32.03 $\pm$ 2.52 & 8215 \\
                         & 2020-11-12.09 &                   & & 32.94 $\pm$ 2.22 & 10178 \\
                         & 2020-11-12.42 &                   & & 32.11 $\pm$ 2.14 & 11358 \\
                         & 2020-11-12.75 &                   & & 45.70 $\pm$ 17.31 & 259 \\
    T04\_066T01\_9000004772 & 2021-11-18.97 & 21.25 $\pm$ 0.12 & 2988 & -- & -- \\
    T04\_072T01\_9000004780 & 2021-11-23.22 & \multirow{9}{*}{22.96 $\pm$ 0.09} & \multirow{9}{*}{25506} & 23.63 $\pm$ 3.06 & 4556 \\
                             & 2021-11-23.41 &          & &  31.22 $\pm$ 2.31 & 8985 \\
                             & 2021-11-23.74 &          & &  26.20 $\pm$ 2.35 & 7802 \\
                             & 2021-11-24.07 &          & &  31.83 $\pm$ 3.27 & 4363 \\
                             & 2021-11-24.41 &          & &  31.79 $\pm$ 2.36 & 9323 \\
                             & 2021-11-24.74 &          & &  34.09 $\pm$ 2.72 & 7314 \\
                             & 2021-11-25.07 &          & &  24.02 $\pm$ 3.39 & 3863 \\
                             & 2021-11-25.41	&          & &  25.80 $\pm$	2.39 & 8201 \\
                             & 2021-11-25.74	&          & &  33.51 $\pm$	4.41 & 2637 \\
    T05\_058T01\_9000005414 & 2022-12-07.23 & 21.38 $\pm$ 0.07 & 7872 & -- & -- \\
    \cmidrule(lr){1-1} \cmidrule(lr){2-2} \cmidrule(lr){3-4} \cmidrule(lr){5-6}
    \textit{Swift} & (UT) & UVOT uvw2 & (s) & XRT $(0.3-1.5)$ keV & (s) \\ 
    \cmidrule(lr){1-1} \cmidrule(lr){2-2} \cmidrule(lr){3-4} \cmidrule(lr){5-6}
    00010498001 & 2018-01-01.22 & 18.87 $\pm$ 0.05 & 575 & $< 39.4$ & 986 \\ 
    00010498002 & 2018-01-02.36 & 19.59 $\pm$ 0.06 & 979 & $< 38.1$ & 991 \\ 
    00010498003 & 2018-01-03.81 & 20.76 $\pm$ 0.16 & 754 & $< 55.9$ & 682 \\ 
    00010498004 & 2018-01-04.49 & 20.27 $\pm$ 0.09 & 1249& $< 30.2$ & 1266 \\ 
    00010498005 & 2018-01-05.48 & 20.69 $\pm$ 0.06 & 1021& 3.6 $\pm$ 1.0  & 5197 \\ 
    00010498006 & 2018-01-05.94 & 20.90 $\pm$ 0.07 & 937 & 11.8 $\pm$ 1.6 & 5317 \\ 
    00010498007 & 2018-01-07.14 & 21.10 $\pm$ 0.08 & 1614& 12.1 $\pm$ 1.7 & 5202 \\ 
    00010498008 & 2018-01-08.07 & 21.00 $\pm$ 0.08 & 1637& 13.8 $\pm$ 2.0 & 4155 \\ 
    
    \bottomrule
    \end{tabular}} 
    \end{table}

Photometric studies were done using images obtained in the ultraviolet (UV) bands from the following two telescopes.

\subsubsection{Swift UVOT}
High cadence UV imaging data of M31N~2008-12a are available from the \textit{Swift} \citep{Geh04} archive \footnote{\href{https://www.swift.ac.uk/index.php}{https://www.swift.ac.uk/index.php}}. The nova has been monitored by UVOT since 2013 during each eruption. The log of \textit{Swift} observations between $2017-2022$ are summarised in Table~\ref{swift_tables}. We have used the uvw2 (1928$\pm$657~\AA) archival data in this study. The \texttt{uvot} task in \texttt{HEASOFT (v6.29)} was used to extract the magnitudes from a source region of radius 5$''$ after background subtraction. Since the field is crowded in $uvw2$ filter, a source-free 10$''$ radius circle, 80$''$ southwards of the object, was chosen to estimate the background. The calibration assumes the UVOT photometric (AB) system (\citealt{poo08} and \citealt{bre11}) and is not corrected for extinction.

\subsubsection{AstroSat UVIT}
\textit{AstroSat} \citep{KPS14} is a space-based telescope with the Ultraviolet Imaging Telescope (UVIT) as one of its instruments. UVIT observed M31N~2008-12a during its $2019-2022$ eruptions in F148W (1481$\pm$500~\AA) filter (see Table~\ref{astrosat_table}). The level 1 UVIT data was downloaded from the Indian Space Science Data Center (ISSDC) \footnote{\href{https://astrobrowse.issdc.gov.in/astro_archive/archive/Home.jsp}{https://astrobrowse.issdc.gov.in/astro\_archive/archive/Home.jsp}} and reduced using \texttt{CCDLAB} following standard routines presented in \citet{postma21}. The orbit-wise images were registered and merged to obtain a single image with a high SNR on which astrometry was performed. The average PSF size in UVIT images was $\sim 1.5''$ across all epochs. We performed PSF photometry with an aperture correction term derived from ``good stars'' to account for the broad PSF wings in UVIT images. The zero points for photometric calibrations in the AB system were adopted from \citet{tandon20} and have not been corrected for extinction.

\subsection{X-ray}
Both \textit{Swift} and \textit{AstroSat} observe simultaneously in the UV and X-ray wavelengths. Soft X-ray observations from both facilities were used to study the eruptions during the SSS phase.

\subsubsection{Swift XRT}
{\it{Swift}} X-Ray Telescope \citep[XRT;][]{burrows05} data were downloaded from the {\it{Swift}} archive, with Observation IDs same as that of the UVOT data (Table~\ref{swift_tables}). For the analysis, \texttt{HEASOFT (v6.29)} with \texttt{XIMAGE (v4.5.1)} and \texttt{XSELECT (v2.5b)} were used following the guidelines summarised by UKSSDC\footnote{\href{https://www.swift.ac.uk/analysis/}{https://www.swift.ac.uk/analysis/}}. 
XRT count rates were determined using both \texttt{XIMAGE} and \texttt{XSELECT} tools provided by \texttt{HEASARC}. Both results followed the same trend and were within 1$\sigma$ errors of each other. We have presented the results from the \texttt{XIMAGE sosta/optimize} analysis as it corrects the counts for vignetting, dead time loss, background subtraction and the PSF of the instrument. We used \texttt{XSELECT} to extract the spectra for each snapshot. ARF files were generated from the exposure maps, while RMF files were taken from the calibration database. Spectral analysis was performed in \texttt{XSPEC (v12.12.0)} assuming Poisson statistics (\texttt{cstat}) due to low counts. We used ISM abundances given in \citet{wilms20} and the Tübingen–Boulder (\texttt{tbabs}) ISM absorption model to account for the intervening medium.

\subsubsection{AstroSat SXT}
AstroSat Soft X-ray telescope SXT \citep{KPS17}, placed in parallel alongside UVIT, is capable of observing in the $\rm 0.3~-~8.0~keV$ range simultaneously with UVIT (Table~\ref{astrosat_table}). SXT observed the SSS phase of M31N~2008-12a during the 2020 and 2021 eruptions. Level 2 data was downloaded from ISSDC, and the cleaned event files were merged using \texttt{SXTTools} in \texttt{Julia}\footnote{\href{http://astrosat-ssc.iucaa.in/sxtData}{http://astrosat-ssc.iucaa.in/sxtData}}. The source region was chosen as a circle of radius $7'$, smaller than the usual SXT  PSF of $\sim 12'$ so to avoid contamination in the crowded M31 field. We set the bin size to 8~hours and energy range to $0.3 - 2.0$~keV in \texttt{XSELECT} (\texttt{XIMAGE} was avoided due to compatibility issues) to attain an adequate SNR for light curve analysis. SXT spectra were extracted using \texttt{XSPEC} from the merged SXT cleaned event files. A new ARF file was generated corresponding to the smaller source extraction region for analysis.

\subsection{Epoch of eruptions}

For a very fast RNe, like M31N~2008-12a, a tight constraint on the eruption time is useful for generating light curve models (\S \ref{sec:lc model}) and studying its recurrence nature (\S \ref{sec:recurrence_nature}). Hence, we estimate the epochs of eruption based on available detection and pre-discovery magnitudes and non-detection upper limits for all eruptions since 2017. Even though the exact time of eruption is uncertain, it can be well approximated by the mid-point of first detection and last non-detection in each year. Amateur astronomers' interest in M31 and the increase in survey telescopes over the past decade have made it possible to constrain the eruption date to well within a day. The uncertainty in the eruption dates spans between the first detection and the last non-detection. 

The 2017 eruption was discovered just in time to be called the ``2017 eruption'' on Dec~31.77~UT by \cite{boyd17}. \cite{dar17c} reported spectroscopic confirmations on the same day. The last non-detection was on Dec 31.38 UT at an upper limit of $\rm m_{clear}=19~mags$ \citep{nai18}. 

The 2018 eruption was discovered on Nov~06.80~UT at a magnitude of $19.15 \pm 0.05$ by \cite{Dar18a} and confirmed spectroscopically by \cite{Dar18b} on the same day. \cite{Tan18} reported the last non-detection at $ >21.20$~mag on Nov 06.54 UT in clear filter. 

The 2019 eruption was detected on Nov~06.71~UT by \cite{Oks19} at 19.40~mags. The first spectrum taken on Nov~06.83~UT \citep{dar19c} confirmed the recurrence of M31N~2008-12a. The last non-detection information was not publicly available for 2019, so we adopted the eruption date provided in \cite{dar19b}. 

\cite{dar20} discovered the 2020 eruption on Oct~30.89~UT. The nova was, however, also detected in images captured 90~minutes before the discovery \citep{gal20}, and we use the pre-discovery detection ($\rm m_g=18.74$) and non-detection ($\rm m_g > 19.4$) to constrain the eruption time. The discovery was spectroscopically confirmed on the next day \citep{dar20c}. 

The 2021 eruption was discovered on Nov~14.38~UT by \cite{ita21} and was spectroscopically confirmed by \cite{wagner21}. \cite{tan21} gave pre-detection upper-limits at $\rm m_{clear}>19.00~mags$ on Nov~13.96~UT. 

The 2022 eruption was discovered on Dec~2.83~UT by \cite{per22}. It was undetected until Dec~02.61~UT at $\rm m_L > 19.60~mags$ \citep{Sha22}. Spectra taken on Dec~3.84~UT confirmed the source to be a recurrence of the nova \citep{dar22a}, its 15th successive eruption in as many years. 

The estimated eruption dates for 2017-2023 are presented in Table~\ref{tab:eruption_all} together with those of the previous ones. 


\section{UV and optical light curve} \label{sec3}

\subsection{Light curve evolution}
\label{subsec:lc_evolve}

\begin{figure}
    \includegraphics[width=0.48\textwidth]{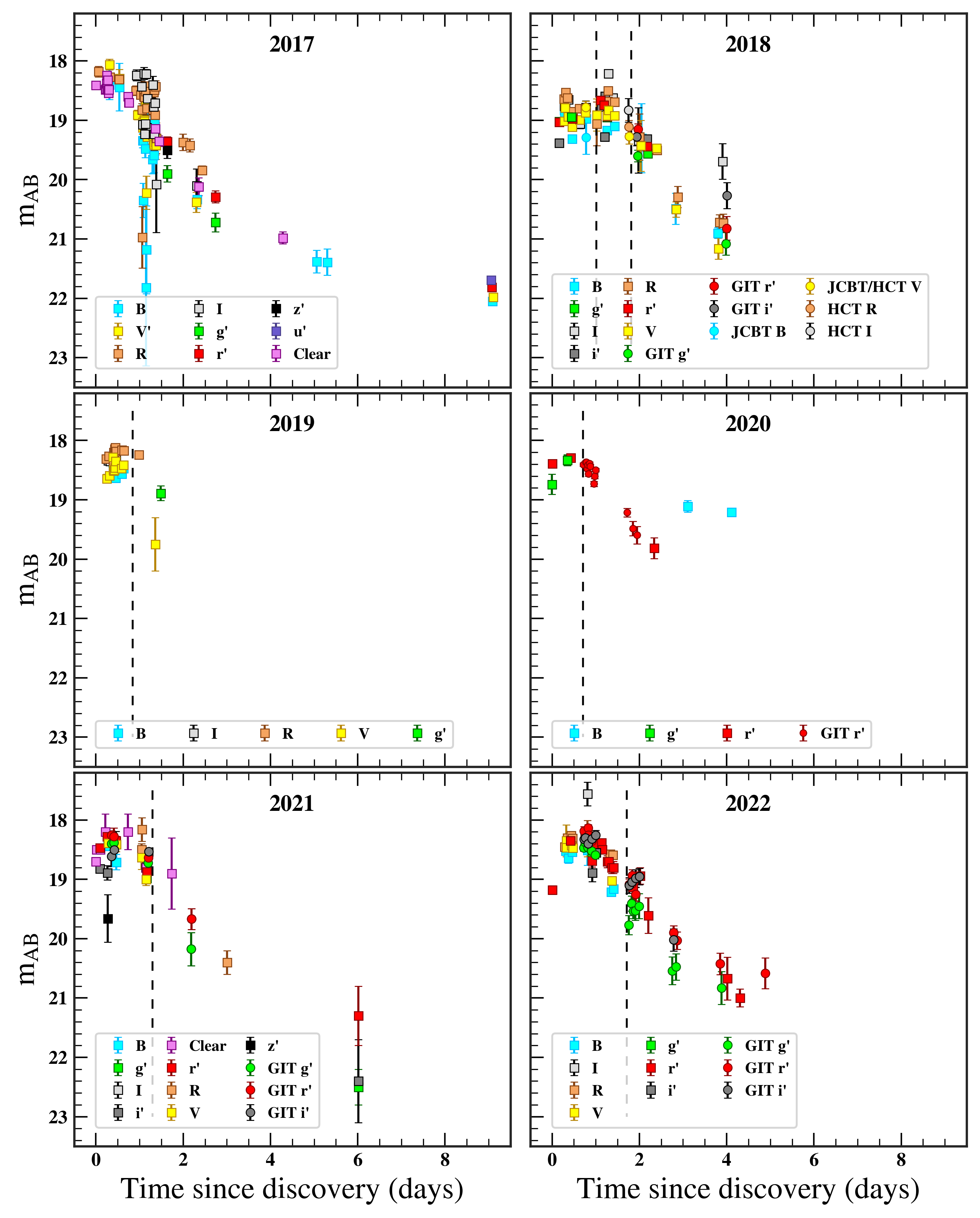}
    \caption{Optical light curves of M31N~2008-12a for $2017-2022$ eruptions. GIT, HCT, and JCBT observations are plotted with publicly available data. Vertical dashed lines in each panel mark the epochs of spectroscopic observations.}
    \label{lc}
\end{figure}

The optical light curves of the $2017-2022$ eruptions, based on our observations and publicly available data, are shown in Figure~\ref{lc}. The light curves indicate a  rapid rise to the peak magnitude in $<1$ day from discovery, followed by a fast decline with $t_2 \approx 2-4$~days in $g'r'i'$ bands. A brief description of the optical light curve for each eruption during $2017-2022$ is provided below.

The 2017 light curves show a rapid decline in the first 4 days, followed by a slow decline in all the bands. The decline rate is marginally faster in the $r'$ band ($\rm 0.84\pm 0.12~mag~day^{-1}$) compared to the $g'$ band ($\rm 0.74\pm 0.19~mag~day^{-1}$). However, the decline rates in $g'$ and $r'$ were measured using only two data points and are within error bars of each other. 

The maximum phase during the 2018 eruption appears to be broader. Following the initial rise, the magnitude declined by $\sim~0.4$~mag in the $BVRI$ filters, and after a brief halt for about 0.3~days at this level, a marginal increase in the brightness by $0.2-0.3$~mag lasting $\sim 0.7$~days can be seen. We also note that the peak magnitude observed in 2018 in $R$~(18.50~mag) is fainter than the peak $R$ or $r$ magnitudes in the other years. In 2018, the initial decline in the $r'$ ($\rm 0.68\pm 0.03~mag~day^{-1}$) and $i'$ ($\rm 0.53\pm 0.04~mag~day^{-1}$) bands was slower by $0.1-0.2$~mag~per~day compared to other years. Further, the decline rates of $r'$ and $i'$ are slower than $g'$ band ($\rm 0.84\pm 0.02~mag~day^{-1}$) in 2018.  

The 2019 data set is sparse and restricted to the initial rise and maximum phases. The nova is brighter in $R$ compared to $BV$ bands during the rise and the peak. It rises about 0.4~mag in all the bands in about 0.5~days from discovery. The peak $R$ magnitude reached in 2019 is $\rm m_R=18.12$, which is higher than most other eruptions.

The $g'$ band traces the rise of the 2020 light curve at $\rm \sim 1.2~mag~day^{-1}$ while the $r'$ band traces the smooth decline from the peak at $\rm 0.89\pm 0.05~mag~day^{-1}$ for 2~days. Limited data only restricts the light curve analysis to the $r'$ band.

The 2021 eruption was caught almost a day before it reached its peak. The rise was sharper in the $i'$ band compared to the $g'r'$ bands. It declined rapidly in the $g'$ band at $\rm 1.25\pm 0.20~mag~day^{-1}$ but relatively slowly in the $r'$ band at $\rm 0.88\pm 0.13~mag~day^{-1}$ for the first $\rm3-4~days$. The decline rate then slowed with significant enhancement in the $r'$ band flux around 6~days after the eruption. 

The 2022 eruption light curve was similar to previous years. The rise is well captured in the $r'$ band with a rate of $\sim$~2~mag~day$^{-1}$, the fastest in the last six years. It then declined speedily in the $g'$ band ($\rm 1.01~\pm~0.05~mag~day^{-1}$) but relatively slowly in $r'$ band ($\rm 0.79~\pm~0.03~mag~day^{-1}$) and even more slowly in the $i'$ band ($\rm 0.69\pm 0.06~mag~day^{-1}$).

\begin{figure}
    \centering
    \includegraphics[width=0.48\textwidth]{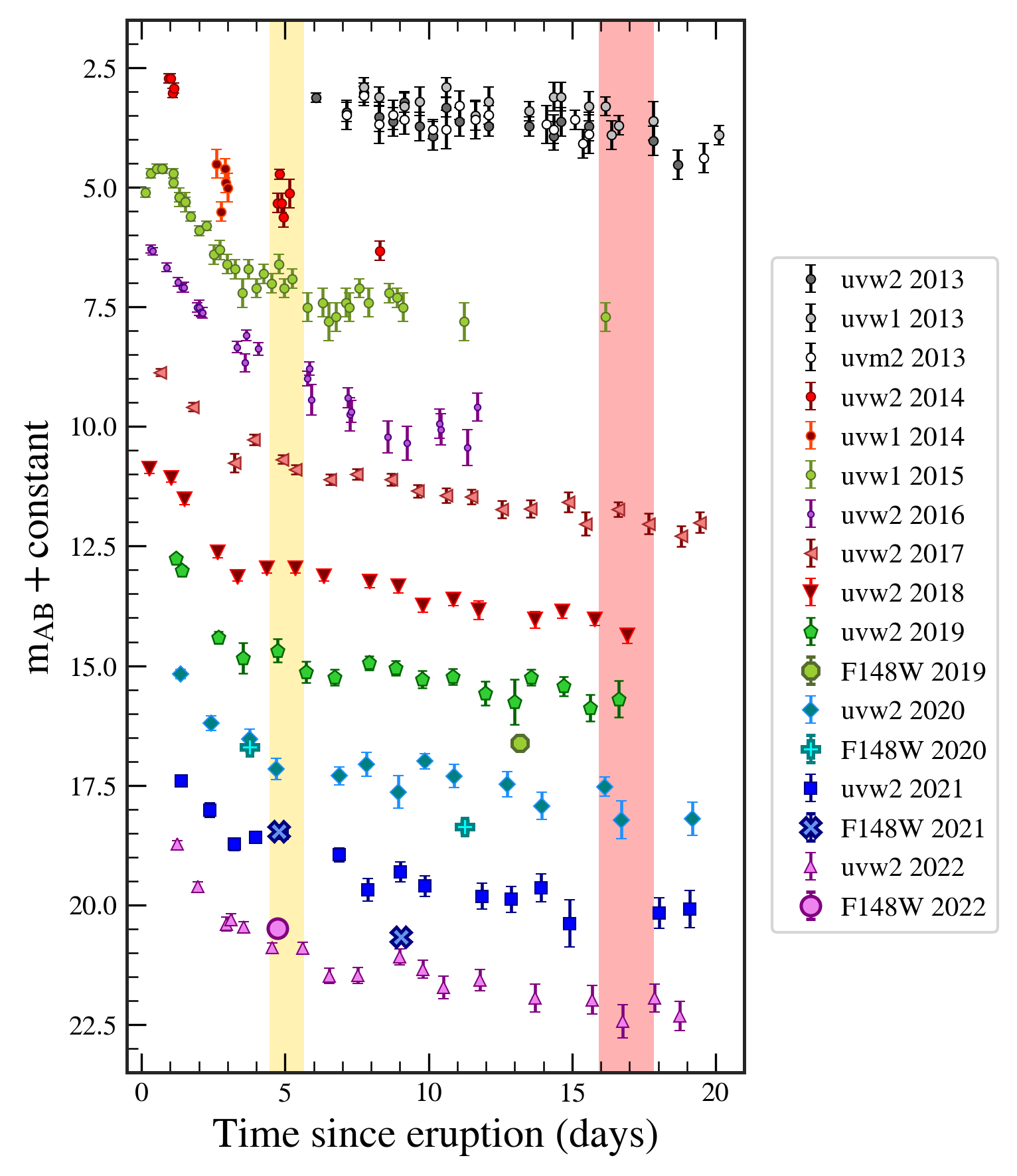}
    \caption{UVOT light curves of M31N~2008-12a shown for $2013-2022$ eruptions. UVIT F148W data are overplotted. SSS $\rm t_{on}$ and $\rm t_{off}$ times are highlighted in yellow and red respectively.}
    \label{uv}
\end{figure}

Generally, the nova declines fastest in the $g'$ band and comparatively slower in the redder bands. \cite{Dar16a} combined the 2013, 2014, and 2015 eruptions and found the decline rate to be fastest in the $V$ band at $1.21~mag~day^{-1}$ during the initial decline phase. The evolution of the optical light curve during the later phases is unavailable as the nova fades beyond the detection limit of the $1-2~m$ class ground-based telescopes generally used for follow-up observations.

The UV light curves (Figure~\ref{uv}) obtained from space-based telescopes have a wider time coverage ($\sim$~20~days) and show a linear decline in the $2017-2022$ $uvw2$ magnitudes from day 0 to 3 since the eruption. The plateau phase begins with a re-brightening at $\sim~4$~days since the eruption, which is also coincident with the SSS turn-on time (yellow shaded region in Figure~\ref{uv}). The evolution during this phase indicates a gradual decline of $\sim$~0.15~mag~day$^{-1}$. However, this decline is not smooth but accompanied by small undulations. Some P-type galactic recurrent novae, such as T~Pyx, RS~Oph and U~Sco, show variability in the $V$-band during their plateau phase \citep{str10}. Since the optical photometry during this phase is insufficient, we are unable to comment on the presence of a similar variability in the optical bands in the case of M31 2008-12a. We encourage continuous and deep optical monitoring during the plateau phase in future eruptions.

The F148W data indicate a brightness similar to $uvw2$ during the initial decline phase, but becomes fainter than $uvw2$ by $>0.5$ mag during the SSS phase. 

The general trends across all the UV--optical bands in $2017-2022$ eruptions are more or less similar over the last six years and consistent with the previous eruptions (2013: \citealt{Dar14}, 2014: \citealt{dar15}, 2015: \citealt{Dar16a}, 2016: \citealt{Hen18}). However, some deviations of the 2016 light curve were noted and are discussed in \S \ref{discussion}.

\subsection{Colour Evolution}
\begin{figure}
    \centering
    \includegraphics[scale=0.63]{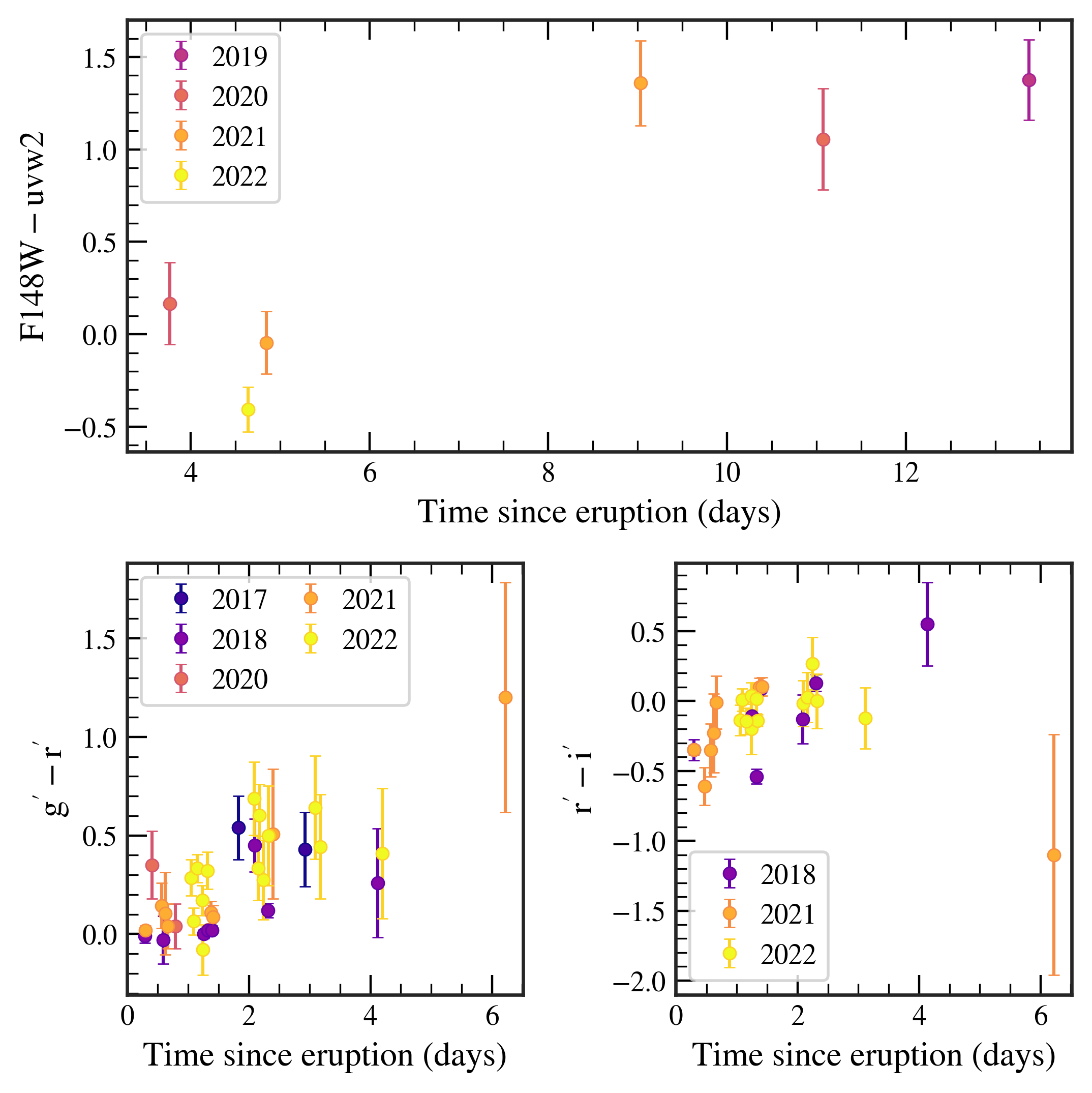}
    \caption{Colour evolution of M31N~2008-12a in UV and optical bands for $2017-2022$ eruptions.}
    \label{fig:color}
\end{figure}

The (F148W$-uvw2$) colour was determined from observations taken on the same day. From Figure~\ref{fig:color}, it is seen that the (F148W$-uvw2$) colour becomes bluer at the onset of the SSS phase but is significantly redder during the SSS phase. 

In the optical bands, we restrict the colour analysis to only the SDSS primed filters to avoid instrumental and/or filter dependencies of the Bessel filters. Near-simultaneous observations in $g'r'$ and $r'i'$ filters for the same eruption were used to estimate the colours. The colours are plotted together as a function of days since the eruption to bring all the outbursts to the same time scale. The ($g'-r'$) colour linearly increases up to day 3 from the eruption and then decreases. This timeline agrees with the initial rise and linear decline phase of the light curve. The ($r'-i'$) colour shows a steep reddening during the rising phase, which then slows down as the nova follows its initial decline. After 3 days from the eruption, the ($g'-r'$) colour becomes bluer while the ($r'-i'$) colour also tends to be bluer, but due to only one data point between day 3 and day 4, we are unable to confirm this. Beyond day 4, the ($r'-i'$) colour becomes redder when the nova enters its plateau phase. A similar trend was also noted by \cite{Dar16a} in their colour plots.

\subsection{Light curve modelling}
\label{sec:lc model}

\begin{figure*}
    \centering
    \includegraphics[width=\textwidth]{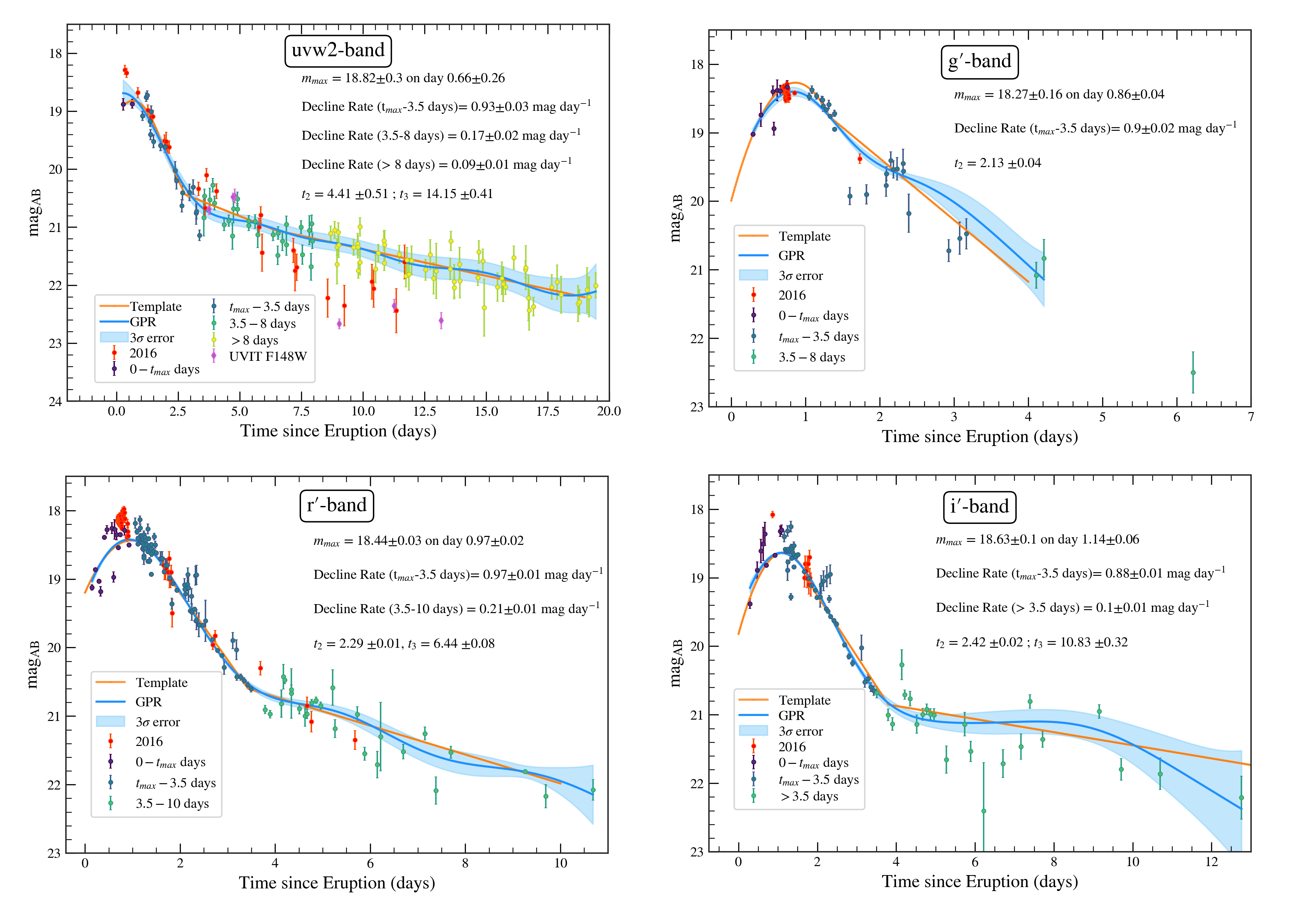}
    \caption{Light curve properties of 2013-2022 (except 2016) eruptions for $uvw2, g', r',$ and $i'$ bands (clockwise from top-left). Outlier data points of 2016 are marked in red. The light curve template, in orange, was generated by stitching the phases in each filter. GP mean and $3 \sigma$ error functions are shown in blue.}
    \label{fig:lc_model}
\end{figure*}

\begin{table*}
    \caption{Multi-eruption light curve model parameters}
    \hspace*{-1cm}
    \begin{tabular}{cccccccc}
    \toprule
    Filter & $\rm t_{max}^a$ & $\rm m_{max}$ & $\rm t_2$ & $\rm t_3$ & \multicolumn{3}{c}{Decline Rates ($\rm mag~day^{-1}$)} \\
     & (days) & (AB) & (days) & (days) & $\rm t_{max}~to~3.5$ & $\rm 3.5~to~\sim 8$ &  $>8$ \\
    \midrule 
    \multicolumn{8}{c}{$2017-2022$} \\
    \midrule   
    \textit{uvw2} & 0.66 $\pm$ 0.26 & 18.82 $\pm$ 0.30 & 4.41 $\pm$ 0.51 & 14.15 $\pm$ 0.41 & 0.93 $\pm$ 0.03 & 0.17 $\pm$ 0.02 & 0.09 $\pm$ 0.01 \\
    \textit{g'} & 0.86 $\pm$ 0.04 & 18.27 $\pm$ 0.16 & 2.13 $\pm$ 0.04 & $\cdots$               & 0.90 $\pm$ 0.02 & $\cdots$ & $\cdots$ \\
    \textit{r'} & 0.86 $\pm$ 0.09 & 18.21 $\pm$ 0.27 & 2.37 $\pm$ 0.05 & 5.87 $\pm$ 5.61 & 0.88 $\pm$ 0.02 & 0.19 $\pm$ 0.18 & $\cdots$ \\
    \textit{i'} & 0.88 $\pm$ 0.12 & 18.30 $\pm$ 0.62 & 3.26 $\pm$ 0.20 & $\cdots$                    & 0.56 $\pm$ 0.04 & $\cdots$ & $\cdots$ \\
    \midrule
    \multicolumn{8}{c}{$2013-2022$ (Excluding 2016) } \\
    \midrule
    \textit{uvw2} & 0.66 $\pm$ 0.26 & 18.82 $\pm$ 0.30 & 4.41 $\pm$ 0.51 & 14.15 $\pm$ 0.41 & 0.93 $\pm$ 0.03 & 0.17 $\pm$ 0.02 & 0.09 $\pm$ 0.01 \\
    \textit{g'} & 0.86 $\pm$ 0.04 & 18.27 $\pm$ 0.16 & 2.13 $\pm$ 0.04 & $\cdots$               & 0.90 $\pm$ 0.02 & $\cdots$ & $\cdots$ \\
    \textit{r'} & 0.97 $\pm$ 0.02 & 18.44 $\pm$ 0.03 & 2.29 $\pm$ 0.01 & 6.44 $\pm$ 0.08 & 0.97 $\pm$ 0.01 & 0.21 $\pm$ 0.02 & $\cdots$ \\
    \textit{i'} & 1.14 $\pm$ 0.06 & 18.63 $\pm$ 0.10 & 2.42 $\pm$ 0.02 & 10.83 $\pm$ 0.32 & 0.88 $\pm$ 0.01 & 0.10 $\pm$ 0.01 & $\cdots$ \\
    \bottomrule
    \end{tabular}
    \\
    \justifying
    $\rm ^a$Days since eruption \\
    \label{tab:lc_model}
\end{table*}

To understand the temporal evolution of the light curves, we model them by breaking them into three phases corresponding to different decline rates. The $uvw2, g', r', \text{ and } i'$ band are used. The eruption times presented in Table~\ref{tab:eruption_all} are used as the reference times, and we measure all other times in \textit{days} with respect to the reference date of each eruption. The three phases considered are
\begin{enumerate}
    \item The rise to peak: from eruption to $\rm t \approx 1.5$. 
    \item The initial steep decline: $\rm t_{max} \leq t \leq 3.5$, where $\rm t_{max}$ is the time of maxima
    \item The slow decline: $\rm t\geq 3.5$ 
\end{enumerate}

Due to extensive coverage in the $uvw2$ filter, we could notice that the rate of decline decreased even further beyond 8 days of eruption. The final phase is thus divided into two segments in $uvw2$. \citet{Dar16a} employed a similar 4-phase division of all light curves to analyse previous eruptions.

First, we generated models for the combined $2017-2022$ eruptions and obtained the light curve properties at different phases given in Table~\ref{tab:lc_model}. Then, we combined the $2013-2015$ eruptions' data (see \S \ref{intro} for references) with those from $2017-2022$ and generated overall light curve models spanning from 2013 to 2022. We note here that the data in $g'$ is sparse as it was not used in most of the observations before 2016. The 2016 data set has been intentionally excluded as an outlier as it deviated significantly from the general trend of other eruptions (plotted in red points in Figure~\ref{fig:lc_model}), especially in the UV light curve. The combined light curve models are presented in Figure~\ref{fig:lc_model} and the light curve parameters are tabulated in Table~\ref{tab:lc_model}. 

Additionally, Gaussian process (GP) regression techniques were employed to fit the entire light curve for each band. The regression results including a $3 \sigma$ error range, are shown in blue in Figure~\ref{fig:lc_model}.

\subsubsection{The rise to peak}

This phase has been modelled with a quadratic function to trace the rise to the peak and the fall just after. Limited availability of $uvw2$ data during this phase led to only partial modelling of the rise in this band. On the other hand, the optical bands have dense coverage of the rise and the peak. The $uvw2$ and $g'$ bands show a smooth rise towards the peak and a smooth decline from the peak. In contrast, the $r'$ and $i'$ bands show a ``cusp'' just before the peak of the modelled light curve is attained. 

On combining the $2013-2015$ light curves with those from $2017-2022$, we clearly see the cusp (Figure~\ref{fig:lc_model}), at least in $r'$ and $i'$ bands, just before the peak is attained. The cusp-like feature is evident in the 2021 and 2022 light curves (Figure~\ref{lc}) as the data points are dense. The 2018 light curve also indicates the presence of the cusp, although with lesser brightness. The 2017 $V$ band and 2019 $R$ band data also hint at the cusp. Observations post-2016 indicate that the cusp is most likely present during all eruptions. The cusp was first noted by \cite{Hen18} in the 2016 eruption in multiple wavebands, who speculated the ``cusp'' could be an isolated event in 2016 (and 2010?), connected to the short SSS phase and long inter-eruption period of the 2016 event. Alternately, as indicated by \cite{Hen18} and \cite{dar20d}, the low cadence observations of the 2013-2015 eruptions during the rising phase could have allowed for this feature to be missed. We suggest this feature is a general trend and not connected to the shorter SSS phase of 2016. However, to confirm it, we encourage very early detection and dense observations during the rise phase in all UVOIR bands of future eruptions. 

The time of maxima was calculated from the quadratic model fits to the data near the peak. The peak magnitudes and the time of the peak are given in Table~\ref{tab:lc_model}. 

\subsubsection{The initial steep decline}
The initial decline is very fast in all the bands. The decline rates during this phase are modelled by a straight line fit from $\rm t_{max}$~to~$\rm t~=~3.5$~days after the eruption. The $uvw2$ decline rate of 0.93~mag~day$^{-1}$ is steeper than the $uvw1$ decline rate (0.78 mag~day$^{-1}$) reported by \cite{Hen18}. In the $2017-2022$ data, we find that the $g'$ band decline rate (0.90~mag~day$^{-1}$) is marginally higher than the $r'$ band (0.88~mag~day$^{-1}$) but when we combine it with the $2013-2015$ data, the $g'$ band decline rate (0.90~mag~day$^{-1}$) is less than the $r'$ band (0.97~mag~day$^{-1}$). \cite{Dar16a} noted that the decline is fastest in $V$ (1.21~mag~day$^{-1}$) whereas the $B$ and $r'$ decline rates (in mag~day$^{-1}$) are 0.99 and 0.97 respectively. The decline rates in this phase are used to derive the $t_2$ times given in Table~\ref{tab:lc_model}. 

\subsubsection{The slow decline}
\label{subsubsec:fd}

The slow decline phase is modelled with a linear fit from day 3.5 onward. This phase consists of the plateau in the light curves which is also coincident with the SSS phase in X-rays. Combining all eruptions from 2013 gives a sufficient number of data points in all the filters for modelling except in $g'$. The decline rate during this phase is low in all the bands (see Table~\ref{tab:lc_model}). The $r'$ band also show scatter during this linear decline, but these jitters are more prominent in the $uvw2$ filter. The $t_3$ time calculated from the straight line fits are 6.17, 10.83, and 14.15~days in $r'i'$ and $uvw2$, respectively. The slowing down of the decline rate can be attributed to the expanding ejecta cooling at $t \geq 4$ days from the eruption. It is also reflected in the colour evolution where after day 4, we see the system become redder (see Figure~\ref{fig:color} and Figure~2 of \citet{Dar16a}). Beyond day 8, in $uvw2$, we see a further decrease in the decline rate and model it with a different slope. Optical photometry is sparse after day 8, but some data points in the $i'$ band are available, though not enough for modelling. The $i'$ band excess ($0.2-0.4$ mags) around day 8 is most notable. This bump is traced by GP regression and is shown in blue in Figure~\ref{fig:lc_model}. 

During this phase, we see a secular trend of decreasing flux with undulations on top of it. This scatter from the smooth decline in $uvw2$ has been discussed in \S \ref{uv_xray_corr}. Towards the end of the final decline phase, when the SSS flux drops to zero at $t>18$, we see a brief period of UV re-brightening before fading away to quiescent.


\section{Optical Spectroscopy}

\subsection{Spectral analysis}

\begin{figure}
    \centering
    \includegraphics[scale=0.45]{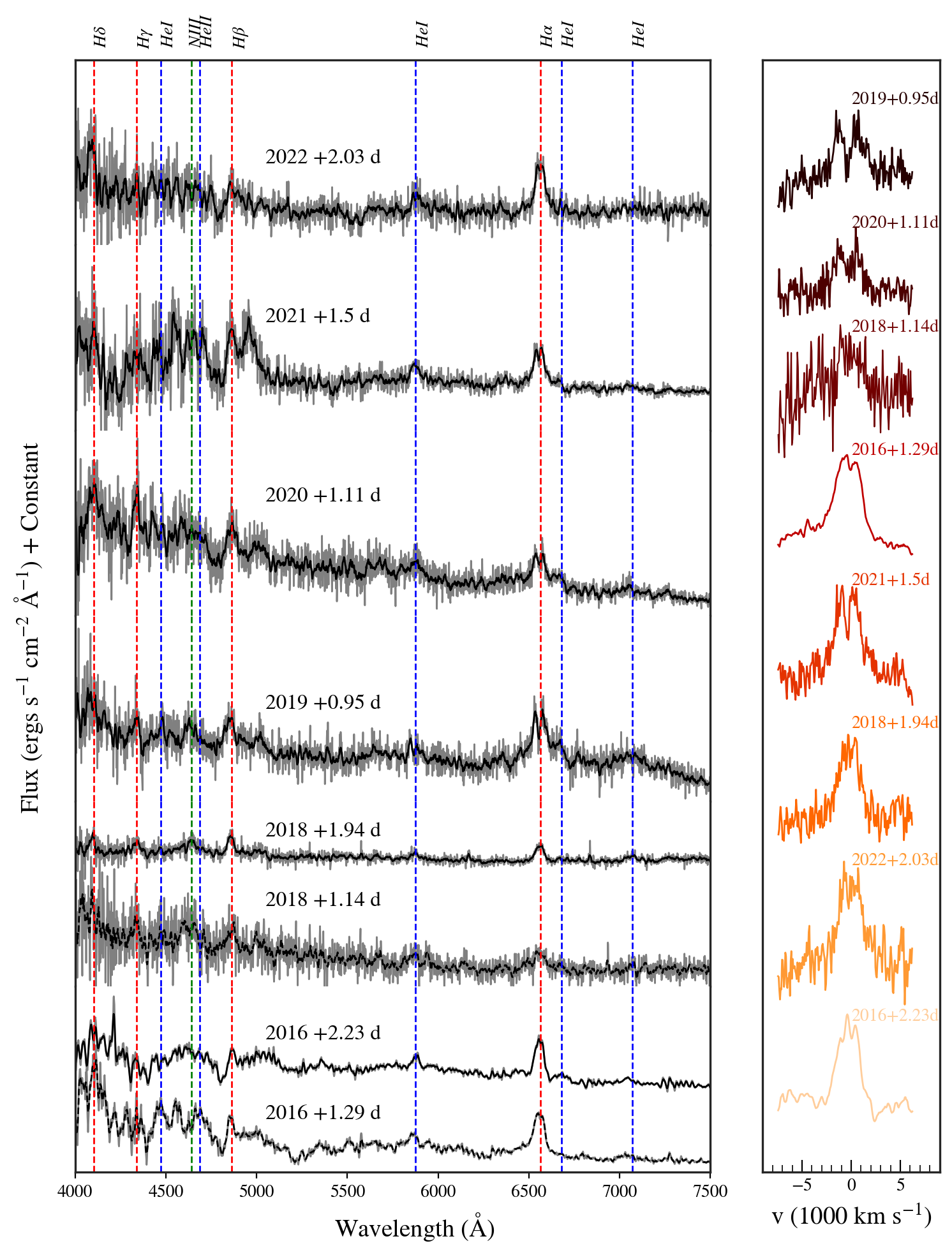}
    \caption{\textbf{Left}: Optical spectra obtained from HCT during 2016, and $2018-2022$ eruptions of M31N~2008-12a. \textbf{Right}: Time evolution of $\rm H\alpha$ morphology from top to bottom.}
    \label{spectra}
\end{figure}

\begin{table*}
\begin{center}
    \caption{Flux and FWHM velocities of identified lines in the spectra}
    \label{tab:obs_spec_v2}
    \resizebox{\linewidth}{!}{%
    \hspace*{-4cm}
    \begin{tabular}{cccccccccccccc}
        \toprule
        \multicolumn{2}{c}{\textbf{Identification}} && \multicolumn{3}{c}{\textbf{2016} (+2.23 d)}  & ~ &  \multicolumn{3}{c}{\textbf{2018} (+1.94 d)}  & ~ & \multicolumn{3}{c}{\textbf{2019} (+0.95 d)}  \\ 
        ~ & ~        && $\lambda$ & Flux $\times 10^{-15}$ & Velocity && $\lambda$ & Flux $\times 10^{-15}$ & Velocity && $\lambda$ & Flux $\times 10^{-15}$ & Velocity \\  
        ~ & ~        && (\AA)  &  (erg cm$\rm^{-2}$ s$\rm^{-1}$)  & ($\rm km~s^{-1}$) & & (\AA)  & (erg cm$\rm^{-2}$ s$\rm^{-1}$) & ($\rm km~s^{-1}$) & & (\AA)  & (erg cm$\rm^{-2}$ s$\rm^{-1}$) & ($\rm km~s^{-1}$) \\  
        \cmidrule{1-2} \cmidrule{4-6} \cmidrule{8-10} \cmidrule{12-14}
        4101 & H I    && $\cdots$ & $\cdots$ & $\cdots$             && 4094.52 & 2.37 $\pm$ 0.60 & 1954 $\pm$ 226 &&  $\cdots$ & $\cdots$ & $\cdots$  \\  
        4340 & H I    && 4333.90 & 2.54 $\pm$ 0.70 & 1952 $\pm$ 271 && 4340.02 & 2.61 $\pm$ 0.41 & 3546 $\pm$ 416 && 4335.70 & 4.08 $\pm$ 0.39 & 2274 $\pm$ 272 \\  
        4471 & He I   && 4489.07 & 0.96 $\pm$ 0.30 & 1637 $\pm$ 258 && 4468.84 & 1.95 $\pm$ 0.47 & 2618 $\pm$ 409 && $\cdots$ & $\cdots$ & $\cdots$ \\  
        4640 & N III  && $\cdots$ & $\cdots$ & $\cdots$             && 4642.11 & 2.01 $\pm$ 0.42 & 2386 $\pm$ 279 && 4632.18 & 6.42 $\pm$ 0.51 & 4113 $\pm$ 233 \\  
        4861 & H I    && 4862.06 & 3.32 $\pm$ 0.91 & 2233 $\pm$ 472 && 4856.94 & 3.35 $\pm$ 0.44 & 2278 $\pm$ 162 && 4850.50 & 6.80 $\pm$ 0.50 & 3121 $\pm$ 158 \\  
        5876 & He I   && 5877.73 & 1.77 $\pm$ 0.37 & 1974 $\pm$ 217 && 5871.05 & 1.22 $\pm$ 0.45 & 2023 $\pm$ 336 && 5861.49 & 1.76 $\pm$ 0.25 & 3092 $\pm$ 292 \\  
        6563 & H I    && 6556.99 & 8.15 $\pm$ 0.39 & 2407 $\pm$ 73  && 6558.89 & 4.42 $\pm$ 0.83 & 2581 $\pm$ 206 && 6559.45 & 23.20 $\pm$ 1.99 & 5099 $\pm$ 180 \\  
        6678 & He I   && 6678.70 & 1.79 $\pm$ 0.50 & 3024 $\pm$ 477 && 6667.21 & 0.57 $\pm$ 0.12 & 1235 $\pm$ 195 && 6671.71 & 3.63 $\pm$ 0.61 & 1921 $\pm$ 172 \\  
        7065 & He I   && 7042.55 & 1.34 $\pm$ 0.08 & 2655 $\pm$ 191 && 7064.87 & 1.79 $\pm$ 0.21 & 3144 $\pm$ 210 && $\cdots$& $\cdots$        & $\cdots$ \\  
        \toprule
        \multicolumn{2}{c}{\textbf{Identification}} & & \multicolumn{3}{c}{\textbf{2020} (+1.11 d)}  & ~ &  \multicolumn{3}{c}{\textbf{2021} (+1.50 d)}  & ~ & \multicolumn{3}{c}{\textbf{2022} (+2.03 d)}  \\ 
        ~ & ~        && $\lambda$ & Flux $\times 10^{-15}$ & Velocity & & $\lambda$ & Flux $\times 10^{-15}$ & Velocity & & $\lambda$ & Flux $\times 10^{-15}$ & Velocity \\  
        ~ & ~        && (\AA)  &  (erg cm$\rm^{-2}$ s$\rm^{-1}$)  & ($\rm km~s^{-1}$) & & (\AA)  & (erg cm$\rm^{-2}$ s$\rm^{-1}$) & ($\rm km~s^{-1}$) & & (\AA)  & (erg cm$\rm^{-2}$ s$\rm^{-1}$) & ($\rm km~s^{-1}$) \\  
        \cmidrule{1-2} \cmidrule{4-6} \cmidrule{8-10} \cmidrule{12-14}
        4101 & H I    && $\cdots$ & $\cdots$ & $\cdots$               && $\cdots$ & $\cdots$ & $\cdots$            && $\cdots$ & $\cdots$ & $\cdots$ \\  
        4340 & H I    && 4336.68 & 6.65 $\pm$ 0.27 & 2000 $\pm$ 52    && $\cdots$ & $\cdots$ & $\cdots$            && $\cdots$ & $\cdots$ & $\cdots$  \\  
        4471 & He I   && $\cdots$ & $\cdots$ & $\cdots$               && $\cdots$ & $\cdots$ & $\cdots$            && $\cdots$ & $\cdots$ & $\cdots$  \\  
        4640 & N III  && $\cdots$ & $\cdots$ & $\cdots$               && $\cdots$ & $\cdots$ & $\cdots$            && $\cdots$ & $\cdots$ & $\cdots$ \\  
        4861 & H I    && 4861.88 & 8.65 $\pm$ 0.55 & 3549 $\pm$ 143   && 4863.05 & 14.8 $\pm$ 2.57 & 3055 $\pm$ 398 && 4870.23 & 6.25 $\pm$ 1.38 & 4015  $\pm$ 472 \\  
        5876 & He I   && 5878.37 & 4.19 $\pm$ 1.13 & 3598 $\pm$ 576   && 5871.49 & 5.20 $\pm$ 0.60 & 2370 $\pm$ 139 && 5880.52 & 4.69 $\pm$ 0.32 & 2839  $\pm$ 122 \\  
        6563 & H I    && 6557.92 & 10.24 $\pm$ 1.47 & 4011 $\pm$ 336  && 6556.97 & 18.80 $\pm$ 0.16 & 3564 $\pm$ 56 && 6562.23 & 15.40 $\pm$ 0.37 & 2924  $\pm$ 53 \\  
        6678 & He I   && 6661.61 & 2.92 $\pm$ 0.10 & 2330 $\pm$ 52    && 6663.74 & 1.73 $\pm$ 0.12 & 1343 $\pm$ 52  && $\cdots$& $\cdots$        &    $\cdots$          \\  
        7065 & He I   && 7044.02 & 1.53 $\pm$ 0.35 & 2386 $\pm$ 447   && 7051.42 & 1.29 $\pm$ 0.10 & 1940 $\pm$ 112 && $\cdots$ & $\cdots$ & $\cdots$ \\  
        \bottomrule
    \end{tabular}}
\end{center}
\end{table*}

The optical spectra with a good SNR are shown in Figure~\ref{spectra} with important emission features marked. Spectra taken on 2021~Nov~14.8~UT and 2022~Dec~3.73~UT were noisy and have not been used for analysis. All the spectra have been dereddened using $E(B - V) = 0.10$ \citep{dar17a}. The spectra were taken within the first three days of the eruption and depict a blue continuum with Hydrogen Balmer and He I~(4471~\AA, 5876~\AA, 6678~\AA) emission lines. Some epochs also show the He II~(4686 \AA) and the N III lines ($\sim$ 4640 \AA). Based on a multi-eruption combined spectrum, \cite{Dar16a} identify several other features in the spectrum. While these features are not seen in the individual spectra presented here, a few faint features could be identified in the merged spectra taken at similar epochs after outbursts. He~I~4922~\AA\ and N~II~5679~\AA\ were detected in the combined spectra of 2018 (1.14d) and 2020 (1.11d), and He~II~4686~\AA, He~I~5016~\AA, N~II~5679 and 6346~\AA, and Raman~O~VI~6830~\AA\ could be identified in the late phase merged spectra of 2018 (1.94d), 2022 (2.03d), and 2016 (2.23d). 

The line fluxes of the emission features clearly identifiable in the individual spectra are listed in Table~\ref{tab:obs_spec_v2}. Also provided in the table are the FWHM velocities obtained from a Gaussian profile fit to the emission lines (using \texttt{IRAF}). The velocities calculated from the widths of the emission lines have been corrected for the instrumental response by de-convolving with the width of night skylines. The initial velocities within 1~day of eruption are as high as 5000~km~s$^{-1}$, typical of very fast novae. These observations are consistent with the previous eruption of M31N~2008-12a \citep{Hen18, Dar16a, dar15}. At around $\gtrsim 1.5$~day after the eruption, the emission line widths narrow to 3000~km~s$^{-1}$. The narrowing of the emission lines could be caused by the expansion and dissipation of the faster moving component \citep{shore96} or by the interaction of the ejecta with the circumbinary material. We find the line velocity decelerating at $\rm v_{exp} \propto t^{-0.27 \pm 0.07}$. This is similar to the estimate provided by \citet{Dar16a} ($\rm v_{exp} \propto t^{-0.28 \pm 0.05}$), who argue that the deceleration is due to the interaction of the ejecta with the circumbinary medium, and that the ejecta is in Phase II of the shocked remnant development \citep{bode85}.

The $\rm H\alpha$ profile, as seen in Figure~\ref{spectra}, indicates that the ejecta geometry is structured and time-dependent. The temporal evolution of the $\rm H\alpha$ line shows a double-peaked structure, prominent in the 2016, 2019, 2020, and 2021 spectra, taken around $\rm0.9-1.5~days$ after the respective eruptions. Around two days after the eruption, the double-peaked profiles give way to a relatively narrow boxy profile.

\subsection{Estimation of physical parameters}
\label{sec:spectra_model}
\begin{figure*}
    \centering
    \includegraphics[width=\textwidth]{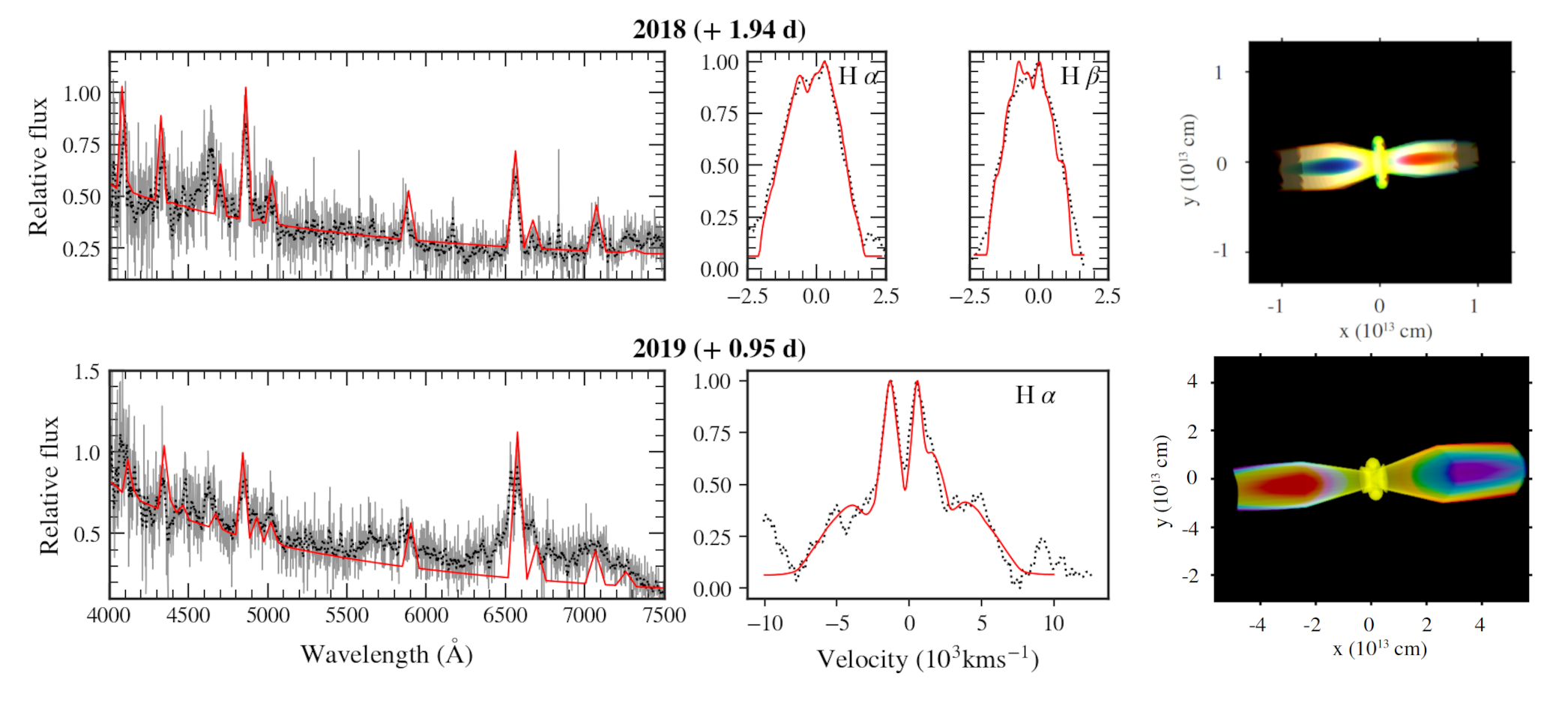}
    \caption{\textbf{Top (from left)}: Best-fit \texttt{Cloudy} modelled spectrum (red) overplotted on the observed spectrum (grey) and smoothed spectrum (black dotted lines) of the 2018 eruption. Best-fit H$\alpha$ and H$\beta$ velocity profiles (red) overplotted on the observed profile (dotted line). Morphology of the ejecta of 2018 eruption obtained from \texttt{Shape} using H$\alpha$ and H$\beta$ velocity profiles. X-axis is the line-of-sight direction, Y-axis is perpendicular to the plane of sky and line-of-sight. Red represents velocity away from us and violet towards us (2500 to -2500~$\rm km~s^{-1}$). \textbf{Bottom (from left)}: Same as top panel, but for 2019 eruption. Velocities are color coded from 6000~$\rm km~s^{-1}$ (red) to -6000~$\rm km~s^{-1}$ (violet). Note that the scales are different in the ejecta morphology plot. }
    \label{fig:spectra_model}
\end{figure*}

To understand the physical conditions in the nova ejecta, the spectral synthesis code \texttt{Cloudy} (\texttt{v17.02}; \citealt{Fer17}) was used to obtain a 1D model using the procedure described in \cite{Pav_thesis}. We generated a 1D model for the best SNR spectra taken on 2018~Nov~8.6~UT and 2019~Nov~7.6~UT. The top left (bottom left) panel of Figure~\ref{fig:spectra_model} shows the 2018 (2019) synthetic spectrum obtained using a two-component (diffuse+clumps) model. 
In the 2018 spectrum, the effective temperature and luminosity of the central ionizing source were found to be $\rm 1.06\times 10^{5}~K$ and $\rm 10^{37}~erg~s^{-1}$ respectively. A clump component and a low-density diffuse component of density $\rm 10^{11}~cm^{-3}$ and $\rm 10^{10}~cm^{-3}$ respectively were used to fit the emission lines in the observed spectrum. The ejected mass and helium abundance from the best-fit modelled spectrum were found to be $\rm 7.21\times 10^{-8}~M_{\odot}$ and $\rm 2.47 \pm 0.11~He_{\odot}$ respectively using the relations given in \citet{Pav19} and references therein. For the 2019 spectrum, the effective temperature and luminosity of the central ionizing source were $\rm 7.20\times 10^{4}~K$ and $\rm 10^{37}~erg~s^{-1}$ respectively. A clump component of $\rm 2.24 \times 10^{10}~cm^{-3}$ and a diffuse component of $\rm 1.26\times 10^8~cm^{-3}$ could generate a synthetic spectrum close to the observed one. The ejected mass and helium abundance, in this case, were found to be $\rm 1.3\times 10^{-8}~M_{\odot}$ and $\rm 3.09 \pm 0.18~He_{\odot}$ respectively.
The ejected mass derived from X-rays (see \S \ref{subsection:SSS phase}) and spectral modelling are similar to that reported for the 2015 eruption by \citet{Dar16a}.
Overabundance of helium has been estimated in other RNe such as RS Oph, V3890 Sgr, T Pyx (see \citealt{gca20} and references therein) and V745 Sco \citep{mondal20}.

It was noted that the two-component model was insufficient to generate the synthetic spectrum with a high $\chi^2$ value. The observed spectrum shows N II lines, which were clearly visible once a third, diffuse, component was introduced to the model. This implies that the N II and He lines are clearly originating from different regions with different physical conditions. However, since the optical spectrum of this extragalactic nova has low SNR, modelling with three components is beyond the scope of this work. With these uncertainties, modelling a high SNR spectrum with similar methods in the upcoming eruptions is recommended.

The H$\alpha$ emission line profile during the 2018 and 2019 eruptions with multiple peaks encouraged us to obtain the morpho-kinematic structure for the ejecta using \texttt{Shape} \citep{ste11}. We carried out the morpho-kinematic analysis of the H$\alpha$ (and H$\beta$ for 2018) velocity profile following the procedure described in \cite{Pav_thesis}.

An asymmetric bipolar structure with bipolar cones and an equatorial ring (Figure~\ref{fig:spectra_model}) with a best-fit inclination angle of 80.75$^\circ$~$\pm$~1.21$^\circ$ could generate the synthetic velocity profile of 2018 spectrum. The extended bipolar component stretched up to 4.52~$\times$~10$^{12}$~cm along the ejecta axis from the centre while the central bipolar cones (opening angle of $\sim$91$^\circ$) extended up to 3.62~$\times$~10$^{11}$~cm. The inner radius of the equatorial ring and radii of the bipolar cones were $\rm 1.27\times~10^{11}$~cm and $5.42\times~10^{11}$~cm, respectively. 
A similar geometry with a best-fit inclination angle of 79.60$^\circ$~$\pm$~1.45$^\circ$ could generate the synthetic H$\alpha$ velocity profile of the 2019 spectrum shown in the bottom panel of Figure~\ref{fig:spectra_model}. The size of the extended bipolar component and the bipolar cone in the central region (opening angle of $\sim$40$^\circ$) were 5.58~$\times$~10$^{13}$~cm and 6.16~$\times$~10$^{12}$~cm along the ejecta axis from the centre respectively. The inner radius of the equatorial ring and the radius of the bipolar cones were $\rm 4.12\times~10^{12}$~cm and $5.27\times~10^{12}$~cm, respectively. It should be noted that the He~I~(6678~\AA) profile is blended with the broad H$\alpha$ profile, and interestingly, the He~I line arises from the inner bipolar cone region.

The outer part of the equatorial ring, the central bipolar cone and the extended bipolar region are discernible in the models shown in Figure~\ref{fig:spectra_model}. The extended bipolar nature suggests a fast-moving polar ejecta along the ejecta axis, i.e. jets, contributing more to the high-velocity hydrogen Balmer emission.

\section{The Super-soft phase in X-ray}

\subsection{X-ray light curve}
\label{subsection:SSS phase}

\begin{figure}
    \centering
    \includegraphics[width=\columnwidth]{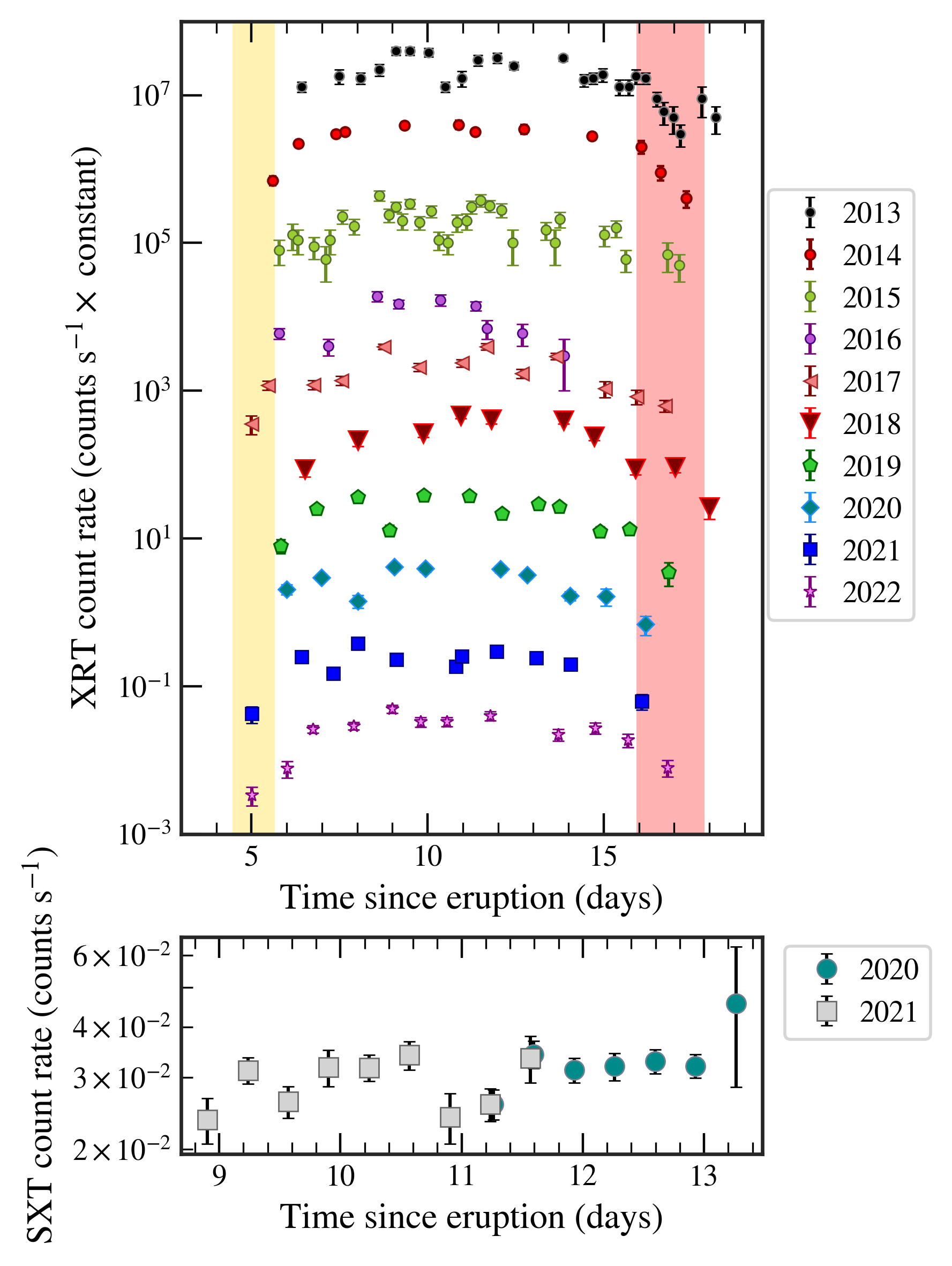}
    \caption{XRT ($2013-2022$) and SXT ($2020-2021$) light curves of M31N~2008-12a during its SSS phase. The SSS turn-on (yellow region) and turn-off (red region) times are also shown.}

    \label{swift}
\end{figure}

Figure~\ref{swift} shows the light curve of supersoft X-ray emission during the $2017-2022$ eruptions. The light curves from the previous eruptions are also shown for comparison. The emergence of the SSS phase is marked by the detections at $\sim~8\times 10^{-3}$~counts~s$^{-1}$, which increases to $(3-4)\times~10^{-2}$~counts~s$^{-1}$ and stays around that level from 8 to 15~days after the eruption. This `peak' of the SSS phase coincides with the UV light curve plateau region. The mean turn-on and turn-off time of the nova estimated from the mid-points of detections and non-detections of $2014-2022$ eruptions are $5.06\pm 0.60$ and $16.89\pm 0.96$~days, respectively from the time of the eruption. The average SSS duration of the nova is $11.83\pm 1.56$~days. Through the rise and during the SSS phase, the X-ray emission is variable, while the decline from the SSS phase is relatively smooth. Multiple ``dips" are seen in the $2017-2022$ XRT light curves.  One around days $6-8$, and an even more noticeable one around days $10-11$. This drop in the count rate is evident in the SXT light curves of $2020-2021$ eruptions (Figure~\ref{swift}). Variability in the X-ray emission during the SSS phase has also been noted in the previous eruptions by \cite{Dar16a} and \cite{Hen18}. The cause of this variability is not yet clear, and further high-cadence observations are required to understand its origin. We also note the unique short and faint nature of the SSS phase in the 2016 eruption compared to other eruptions, which is quite apparent in Figure~\ref{swift}.

\begin{figure}
    \centering
    \includegraphics[width=\columnwidth]{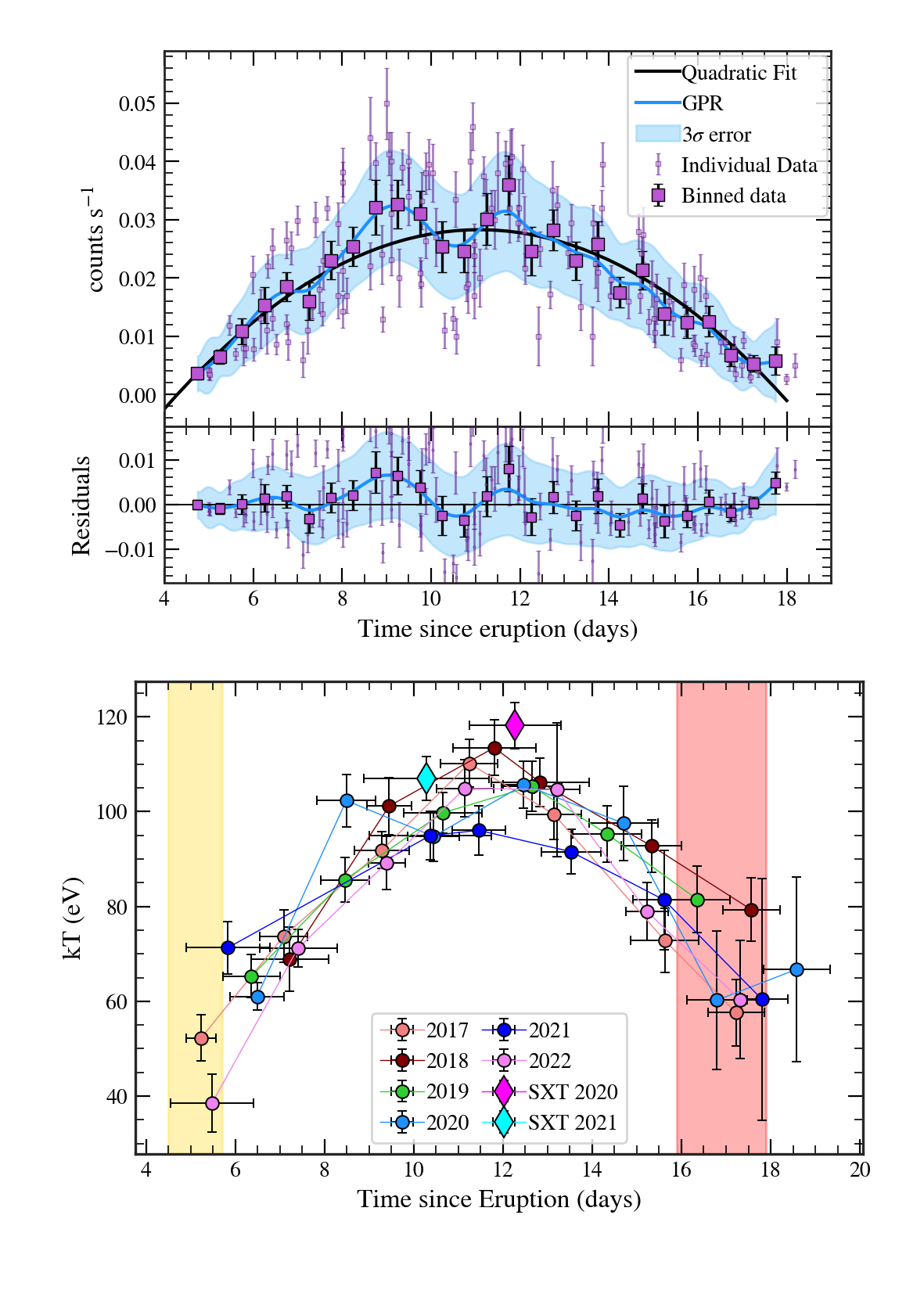}
    \caption{\textbf{Top panel}: Combined XRT data of the SSS phase during 2013-2022 (except 2016) eruptions. Overplotted are data points binned at 0.5 days, its quadratic fit in black, GP regression and its corresponding $3\sigma$ error region in blue. The deviations from the simple quadratic fit are shown below it. \textbf{Bottom panel}: Temperature evolution from XRT data during the SSS phase of $2017-2022$ eruptions. SXT data points for 2020 and 2021 are over-plotted. Mean turn-on and turn-off times are marked in yellow and red, respectively.}
    \label{fig:xray_lc_model}
    \label{fig:SSS_temp_evol}
\end{figure}

To model the rise and decline in soft X-ray flux, we used a simple quadratic function. We included all the observations from 2013, except the peculiar 2016 eruption as its effect was seen most in the SSS phase. We plot all the individual and binned sets in Figure~\ref{fig:xray_lc_model}.  Deviations from the naive quadratic function are evident, especially the peaks at days $9-10$ and $11-12$ and the dips at days $7-8$ and $10-11$. The prominent features between days 8 and 13 are present in the binned and unbinned data, indicating that these variabilities' causes last for more than half a day. The drop and rise of flux between days 10-11 seem to be a general feature of the SSS phase of M31N~2008-12a. Most of the variability is seen up to day 13, whereafter, the decline is relatively smooth.

X-ray studies of M31 novae (\citealt{Hen10, Hen11, Hen14a}) have revealed the correlation of ejecta expansion velocity and the SSS $\rm t_{on}$ time. The ejecta mass was calculated from the turn-on times and the ejecta velocities ($\rm v_{exp}$) using the relation given in \cite{Hen14a}. A $\rm t_{on}$ time of $\sim$5 days with an $\rm v_{exp}$ of $\rm \approx~2000~\pm~200~km~s^{-1}$ around this phase gives an ejecta mass range of $\rm 4.2~\times~10^{-8}~<~M_{ej,H}~/~M_{\odot}~<~10.2~\times~10^{-8}$. These are slightly higher than that calculated from optical spectra in \S \ref{sec:spectra_model} but less than the average mass accreted in a year.

\subsection{X-ray spectroscopy}

\begin{figure}
    \centering
    \includegraphics[scale=0.75]{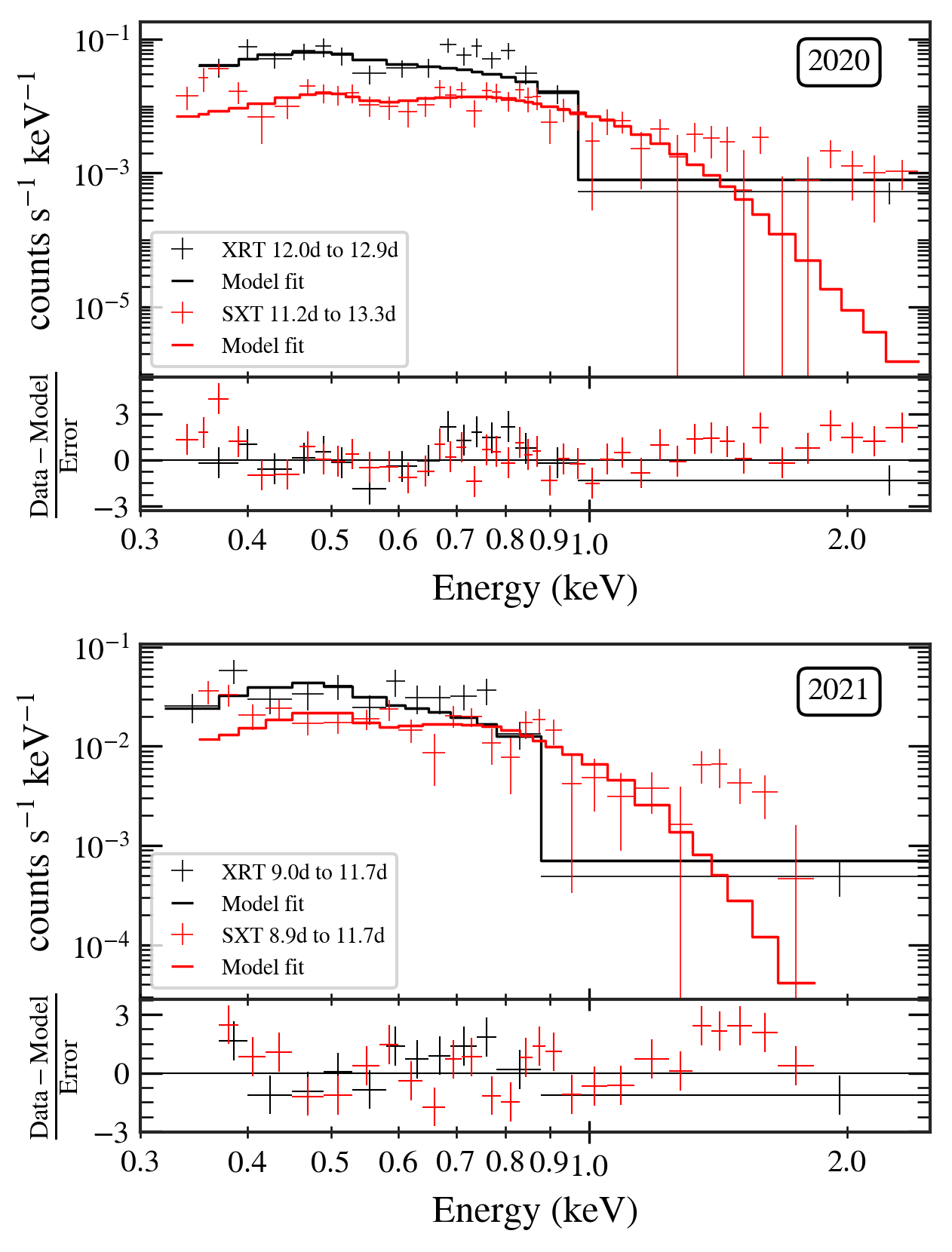}
    \caption{XRT and SXT spectra of 2020 (top panel) and 2021 (bottom panel) eruptions. XRT data nearest to the SXT observation dates have been used for better comparison. The exact observation epochs are given in the legends (in days since the eruption). Spectral fitting involved a single black body with ISM absorption models for both SXT and XRT.}
    \label{fig:xray_spectra}
\end{figure}
The XRT data was also used to extract spectra by merging two data sets obtained on consecutive days to increase the SNR. We fixed the H column density at $\rm 1.4\times 10^{21} cm^{-2}$ \citep{Dar16a} but varied the blackbody temperature and normalization to attain the best-fit values. The time evolution of the SSS temperature is shown in Figure~\ref{fig:SSS_temp_evol} for $2017-2022$ eruptions. Not only does the fluxes peak $10-14$ days after the eruption, but the temperatures also peak, suggesting a correlation between the SSS flux and temperature. In the 2020 eruption, a temperature fluctuation during the maxima can be seen. Such fluctuations have been reported before by \citet{Dar16a}. This pattern is not seen in other eruptions, possibly due to combining data sets of two consecutive days. In \S \ref{subsection:SSS phase}, it was noted that the rise to the maxima in the SSS phase shows variability, whereas the decline was smooth. The temperature evolution in Figure~\ref{fig:SSS_temp_evol} also shows an asymmetry during the rise and decline of the SSS phase. These could be because of two different underlying causes. The increase in flux and temperature is due to the expansion and thinning of the ejecta, probing the deeper and hotter layers towards the WD surface. Whereas during the later stage, when the obscuring material is already dissipated, the decrease in flux and temperature is because of the residual nuclear burning slowing down and eventually stopping.

Spectra extracted from the merged SXT data is shown in Figure~\ref{fig:xray_spectra}. Also shown are the contemporaneous XRT spectra obtained from merged snapshots of two successive days. The data has been restricted to below 2.5~keV for the SXT to avoid background contamination due to its large PSF compared to the XRT. Beyond 1~keV, the flux is too low, owing to the super-soft nature of the source. A faint hard X-ray tail (above 1.5~keV) can be seen in SXT data, but we could not be certain of its origin because of low SNR. The best-fit blackbody temperatures from SXT spectra are slightly higher than the XRT data during similar times (Figure~\ref{fig:SSS_temp_evol}). The modelled flux in the $0.3-2.0$~keV range was similar in the 2020 spectra for both instruments but differed by a factor of 2 (higher in SXT) in the 2021 spectra. As the observations are not continuous and the XRT and SXT epochs do not coincide exactly, the mismatch could be because of the rapid variability seen in flux and temperature during the SSS phase in recurrent novae.

\section{UV -- X-ray correlation?}
\label{uv_xray_corr}
\begin{figure}
    \centering
    \includegraphics[scale=0.65]{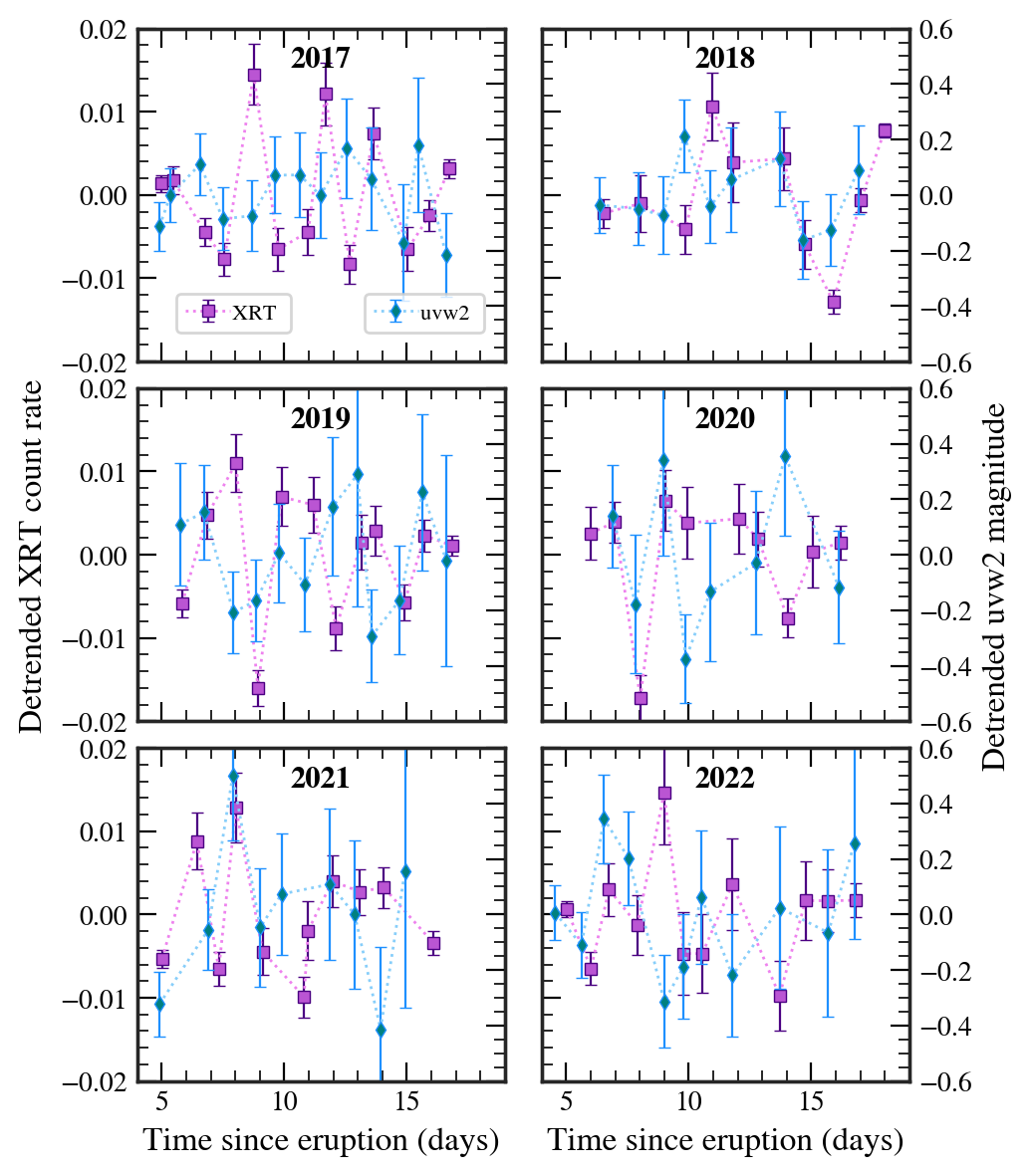}
    \caption{Detrended Swift $uvw2$ and XRT light curves for $2017-2022$ eruptions during the SSS phase.}
    \label{fig:UV_xray_correlation}
\end{figure}

We noticed the 2016 $uvw2$ light curve was ``shorter and less luminous'' compared to the $2017-2022$ $uvw2$ measurements during the SSS phase. \cite{Hen18} had found the same for soft X-rays and reasoned it to be due to a reduced accretion rate prior to the 2016 eruption. They could not comment on the 2016 $uvw2$ measurements because of the unavailability of the $uvw2$ light curve template at that time. 
This motivated us to find the connection between these two wave bands in the super-soft phase. Since both soft X-ray and UV show different trends during the super-soft phase, it was necessary to detrend them. The $uvw2$ light curves were detrended with a linear fit as it followed a linear declining trend during SSS phase, whereas the X-rays were detrended with a quadratic function as it followed a rise and a subsequent fall.
The detrended light curves of each year from 2017 to 2022 are given in Figure~\ref{fig:UV_xray_correlation}. In 2017, we saw the UV and the X-ray fluxes behave inversely between days 7.5 and 13, and a Pearson correlation coefficient (hereafter $r$) value of -0.76 suggests a strong anti-correlation. In 2018, the anti-correlation lasts shorter from day 8.5 to day 11.5 but is stronger with a $r$ value of $-0.86$. 2019, on the other hand, does not show any strong correlation. In 2020 and 2022, there was a strong anti-correlation from day 9.5 to 14.5 when $r$ value was $-0.79$ and $-0.68$, respectively. The 2021 detrended light curves show mild anti-correlation ($r=-0.38$) which is stronger than 2019 but weaker than the other eruptions during the days $9-14$. This anti-correlation seen in most of the eruptions between UV and soft X-ray is strongest from day 8 to day 14 after the eruption, a time corresponding to the maxima of the SSS phase. UV and X-ray flux of most novae have been found to be uncorrelated \citep{Page22}. Nonetheless, \cite{ness09} noted such anti-correlation of UV and X-rays in the detrended light curves for nova V458 Vul, although in the $\rm0.6-10~keV$ X-ray range, just before the start of the SSS phase, and they suggested that it would imply that both the UV and hard X-rays originate from the same region. In HV Cet, the UV and X-ray flux were found to be tied up in phase, which \cite{beard12} argued to be due to the same cause, the orbital period in their case. V603 Aql also showed correlated UV--X-ray emission, though in this case, it was interpreted as due to X-ray illumination \citep{bor03}.
Since the source of soft X-rays during the SSS phase is the nuclear burning on the surface of the WD, the anti-correlation during the SSS peak, in our case, would hint that the UV radiation origin is also close to the surface of the WD. 
It is possible that the accretion disk survives each eruption \citep{dar17a, dar17b}. The surviving partial accretion disk would emit UV radiation. 
The complete reformation of this partial disk could cause the variability seen in both UV and X-ray detrended light curves. The possibility of a wobbly, nascent accretion disk could also cause such a behaviour.


\section{Recurrence period, Accretion rate and WD mass }
\label{sec:recurrence_nature}
M31N~2008-12a has erupted every year since 2008, making it an exceptional case of the only RN observed 15 times and that too consecutively. This section focuses on the trend of the recurrence period and its relation to the accretion rate and the WD mass. 

\subsection{Increasing recurrence period}

\begin{figure*}
    \centering
    \includegraphics[width=\textwidth]{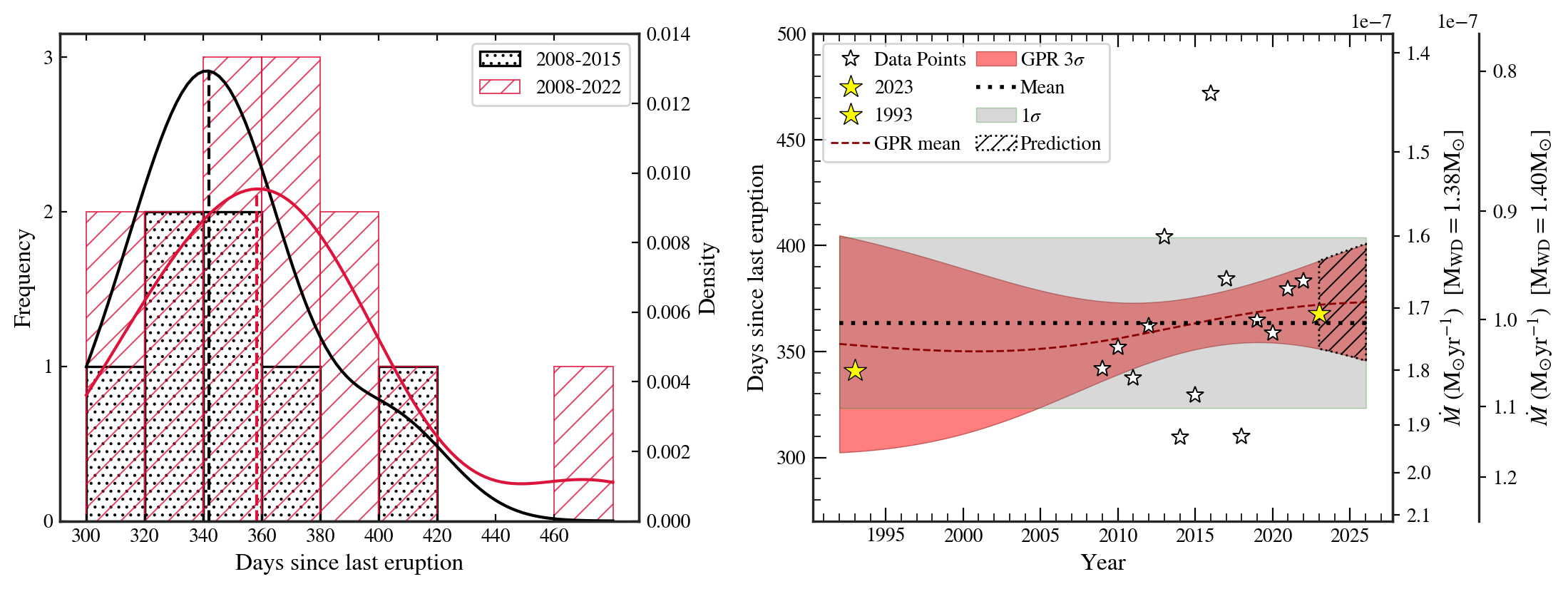}
    \caption{\textbf{Left:} Distribution of the frequency of eruptions as a function of `days since the last eruption' binned at 20~days. Black histogram is for $2008-2015$ eruption data, and black solid line represents the KDE for this data. Red histogram and red solid line represent the same for $2008-2022$ data.
    \textbf{Right:} Eruptions marked with stars. Dashed lines and coloured areas represent the mean and error functions of the two models tracking the recurrence period with time. Shaded region is the projection for the $2023-2025$ eruptions, given we don't see outlier events. Secondary Y-axes represent the accretion rates for $\rm 1.38~M_{\odot}$ and $\rm 1.40~M_{\odot}$ WD corresponding to the recurrence period on the primary y-axis.} 
    \label{fig:erupt_predict}
\end{figure*}

The mean recurrence period was reported to be $\rm P_{\text{rec}}~=~351\pm~13$~days after the 2015 eruption by \cite{Dar16a}, which was updated to $363\pm 52$~days after the outlier 2016 event by \cite{Hen18}. 
Since the 2016 eruption, M31N~2008-12a erupted six more times, and each year, the time gap between two successive eruptions has been more than the mean recurrence period except in 2018 (310.1~days) and 2020 (355.9~days). 
As of the 2022 eruption, the mean recurrence period is $\rm P_{\text{rec}}=363.6$~days with a standard deviation of 40.3~days (Figure~\ref{fig:erupt_predict}, right panel). On the other hand, the median recurrence period has increased from 347~days in 2016 \citep{Hen18} to 360.5~days in 2022. 
Figure~\ref{fig:erupt_predict} shows two different sets of histograms, along with their kernel density estimates (KDE), for the recurrence periods between $2008-2015$ and $2008-2022$. On considering up to the 2015 eruption (grey histogram in Figure~\ref{fig:erupt_predict}), the mode of the recurrence period is 340 days and the KDE peaks at 341.28~days (FWHM of 67.27~days). But on incorporating all the eruption information till 2022 (red histogram in Figure~\ref{fig:erupt_predict}), the mode of recurrence period shifts to 360~days with the KDE peaking at 358.18~days (FWHM of 96.36~days). Over the last seven years, we see a clear increasing trend in the recurrence period.  

To further investigate this matter, we plotted the period as a function of the eruption year in the right panel of Figure~\ref{fig:erupt_predict}. We used the GP regression technique to extract the trend in the data set and associate errors with it. The data was modelled using a Matern kernel with a typical length scale of 15~years and an amplitude equal to the median of the recurrence period. We tested our model by applying it to the eruption dates of $2008-2021$, and it could predict the 2022 eruption date within $3\sigma$ error limits. The 2022 data was subsequently included in the training set to project the upcoming eruptions. Since there is a gap of around 15 years between 1993 and 2008, we did not include the 1993 data point for our modelling, but an extrapolation of our model does seem to incorporate the 1993 eruption within error limits. For comparison, the mean and 1$\sigma$ of the constant recurrence period model have also been shown in Figure~\ref{fig:erupt_predict}. Both the models could reasonably anticipate the 2023 eruption. However, the GPR model can pick up any underlying data trends and better constrain the change in accretion rate.

We emphasize here that our model is restricted to only the `usual' eruptions that follow the trend. Anticipation of `outlier' events, such as the 2016 eruption, is not feasible.

\subsection{Estimating the WD mass}

\begin{figure}
    \centering
    \includegraphics[width=\columnwidth]{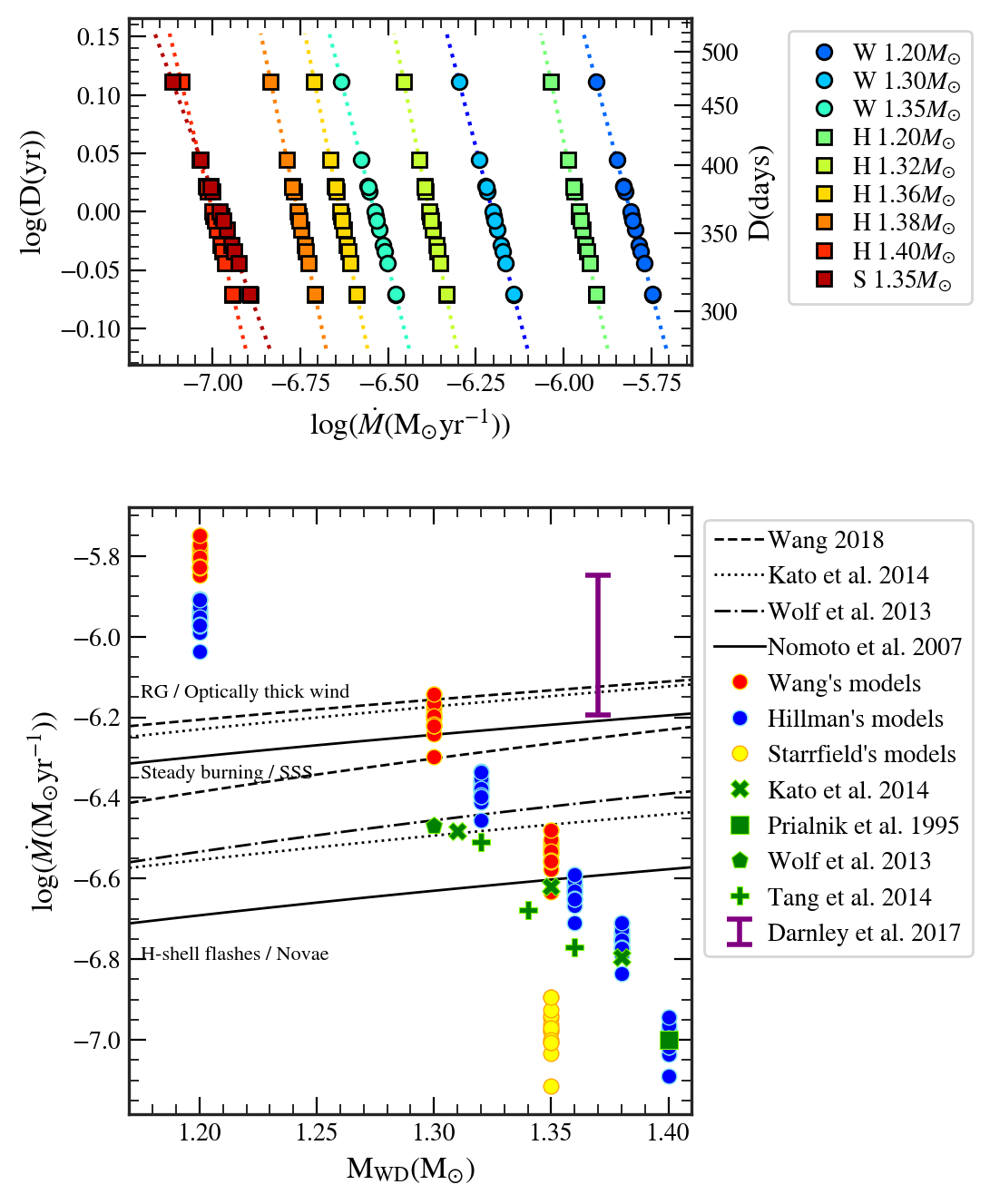}
    \caption{ \textbf{Top}: Observed recurrence period plotted against accretion rate for different WD masses. ``W", ``H" and ``S" in the legend indicate accretion rates derived from \protect \cite{wang18}, \protect \cite{Hill16} and \protect \cite{Starr17} respectively.
    \textbf{Bottom}: $\rm M_{WD}-$$\dot{M}$ parameter space for M31N~2008-12a. Accretion rates from the top panel are shown in red (Wang's), blue (Hillman's) and yellow (Starrfield's). Overplotted in green markers are for the one-year recurrence period taken from the literature. Stable and critical accretion rate limits from four studies are shown as horizontal tracks. The accretion rate range obtained by \protect \citet{dar17b} during quiescence is shown in purple.} 
    \label{fig:acc_rate_period}
\end{figure}

In a theoretical study to understand the possible mass growth of a WD accreting matter from a non-degenerate companion, \cite{Hill16} have explored a range of accretion rates and derived limits on the accretion rate and on the initial mass that will allow a WD to reach the Chandrasekhar limit. 
Adopting their relation between $D$ (period) and $\dot{M}$ (accretion rate) for hydrogen accretion cases, $\log~\dot{M}~=~\rm-A~\log~D~-~B$, we estimate the accretion rates in the last 15 years for the M31 RN eruptions. Here, the coefficients A and B depend on the WD mass. For each value of $\rm M_{WD}$, A and B were determined by fitting a linear function to the parameter space of $\rm\log~D$ and $\log~\dot{M}$. The accretion rates of WD masses between $\rm1.20~M_{\odot}$ and $\rm1.40~M_{\odot}$ corresponding to the periods of M31N~2008-12a are shown in the top panel of Figure~\ref{fig:acc_rate_period}.

\cite{wang18} have used the Modules for Experiments in Stellar Astrophysics (\texttt{MESA}) to model the binary evolution of WD accreting H-rich material from a companion for a range of WD masses and accretion rates.

The composition of the accreted material was fixed at H:He:Metals $\equiv$ 70:28:2. We use their results for the massive WD cases ($\rm1.20-1.35~M_{\odot}$), obtain a best-fit power-law relation between the accretion rate and the period and employ the same to infer the accretion rates for each cycle of M31N~2008-12a. These are also shown in the top panel of Figure~\ref{fig:acc_rate_period}.

We plot the accretion rates thus obtained corresponding to the periods of M31N~2008-12a in the last 15 years in the $\rm M_{WD}-$$\dot{M}$ parameter space in the bottom panel of Figure~\ref{fig:acc_rate_period}. For comparison, we over-plot the results from \cite{kato14}, \cite{pri95}, \cite{wolf13}, and \cite{tan14}, who predicted the WD mass and accretion rates for a recurrence period of 1 year. 

In the $\rm M_{WD}-$$\dot{M}$ plane, when the accretion rate surpasses the critical threshold ($\dot{M}~>~\dot{M}\rm_{cr}$), the WD exhibits a red giant (RG) like behaviour, undergoing surface mass burning at the critical rate, while any excess material is ejected in the form of optically thick winds  \citep{kato94, hachisu96}. Conversely, when the accretion rate falls below the critical threshold but remains above the stable H accretion rate, i.e. $\dot{M}\rm_{cr}$~$~>~\dot{M}~>$~$\dot{M}\rm_{st}$, the burning on the WD's surface remains stable, capable of sustaining itself over an extended period as a supersoft X-ray emitter \citep{kato14}. Below the stability line, i.e. $\dot{M}\rm_{st}$$>\dot{M}$, the accretion rate is insufficient to sustain continuous hydrogen burning. Systems within this parameter range experience ``H-shell flashes" or nova eruptions, a category that includes all RNe, including M31N 2008-12a. 

The limits on $\dot{M}\rm_{cr}$ and $\dot{M}\rm_{st}$ taken from \cite{wang18}, \cite{kato14}, \cite{wolf13}, and \cite{nomo07} are shown in the bottom panel of Figure~\ref{fig:acc_rate_period}. The differences in these limits are primarily because of the different techniques employed in modelling.

We infer from Figure~\ref{fig:acc_rate_period} (bottom panel) that a WD with mass below 1.30~$\rm M_{\odot}$ has a high accretion rate and does not allow the RN phenomenon to occur at the rate of once a year.

We also see that $1.32-1.36~\rm M_{\odot}$ WDs do fall into the ``H-shell flash" region of some of the models, and any WD with $\rm M_{WD}~>~1.36~M_{\odot}$ satisfies the necessary criteria of nova eruption ($\dot{M}$$<\dot{M}\rm_{st}$) for all the models. Thus, the WD mass in M31N~2008-12a is likely to be greater than $\rm1.36~M_{\odot}$, which would, in turn, allow ``H-flash" features at the observed recurrence period of M31N~2008-12a. 

However, it should be emphasized that \cite{Starr17} generated models using \texttt{MESA}, which allow for the stability line (and the critical line) to exist for less massive WDs but not for higher mass WDs ($\sim \rm 1.35~M_{\odot}$ model shown in Figure~\ref{fig:acc_rate_period}). It was shown that such massive WDs would show H-flashes initially, followed by He-flashes, and ultimately grow to $\rm M_{Ch}$. 

Strikingly, none of the models could predict the accretion rate derived from accretion disk modelling discussed in \cite{dar17b}. The accretion rate was found to vary during eruption, SSS, and quiescence phases. In Figure~\ref{fig:acc_rate_period}, we show the range of accretion rate onto the WD during quiescence. It may hint that Starrfield's models, which do not predict any upper limit on stable accretion, are better suited for such massive systems. But at the same time, the accretion rates generated by both types of models (with and without an upper limit) are insufficient to match the ones derived from observations of this exceptional RN.

\section{Discussion}
\label{discussion}

\subsection{More evidence of jets?}
The cusp around maxima in the light curves has been suggested to have different origins. It could be due to shock from a secondary ejection \citep{Kat09}, polar outflow along the line-of-sight, or ejecta-donor interaction \citep{Dar18a}. Observational evidence of broad-winged emission features supports the presence of a fast-moving component in the ejecta. Modelling the H$\alpha$ (and H$\beta$ for 2018) line profiles using \texttt{Shape} could also generate these fast-moving polar ejecta close to the line-of-sight. \cite{dar17a} proposed the presence of these jets using HST data, though they did not confirm it. The photometric and spectroscopic evidence combined with H$\alpha$ profile modelling presented in this work further strengthens the claim of polar jets emanating close to the line-of-sight, indicating a low-inclination angle for this system.  

Optical imaging and spectroscopic modelling of RS Oph revealed the ejecta to be bipolar, possibly associated with jets \citep{bode07, ribeiro09}. RS Oph jets were confirmed in radio wavebands \citep{obrien06, rupen08, soko08, munari22}. High-velocity components of emission lines are standard signatures of jets and were observed in V1494 Aql \citep{iij03}, U Sco \citep{kato03}, V6568 Sgr and YZ Ret \citep{McL21a}. \cite{McL21} pointed out that jets are usually associated with fast novae, which are essentially linked to massive WDs. M31N~2008-12a falls perfectly into these categories. One possible mechanism for jet formation could include bipolar winds during the nova outburst and subsequent mass ejection into an asymmetric medium. This case is particularly interesting for RNe where the asphericity left behind from previous eruptions triggers material to escape through certain channels. Magnetic field lines near the WD and the accretion disk could also lead to the collimation of wind perpendicular to the disk \citep{ogi01}. The accretion disk of M31N 2008-12a is known to be luminous \citep{dar17b}. Such bright disks can also give rise to supersonic winds forming jets \citep{fukue02}. 

\subsection{Light curve}
The optical light curves from 2017 to 2022 are similar. A sharp linear decline is seen from day 1 since the maximum, followed by an approximately flat but jittery plateau and then the final decline ensues. The evolution is similar to the past eruptions and is close to the light curves of P-class recurrent novae \citep{str10}.

The UV peak is observed before the optical peak in all the eruptions. The UV light curve shows a decline from peak magnitude until the onset of the plateau phase, followed by multiple jitters. The UV plateau phase is consistent with the SSS phase's turn-on time. A flat decline follows it and ultimately ends with a brief period of brightening. The 2016 \textit{uvw2} light curve shows considerable deviation from the other eruptions. 

The SSS phase turns out to be similar in all the eruptions except for the 2016 eruption, where it ended as early as $\rm \sim 16 ~ days$ from the eruption. The SSS temperature is strongly correlated to the soft X-ray flux. There is a significant drop in X-ray flux during the $2017-2022$ eruptions around the same time that was noted in the previous eruptions on day 11 since the eruption. The cause of this drop in flux is yet to be explored.

During the slow decline phase, the 2016 $uvw2$ light curve (Figure~\ref{fig:lc_model}) deviated from the general trend. 
Here, we also point out that for the first time in 2016, detailed $uvw2$ observations were conducted, and the light curve was found to be similar to the 2015 $uvw1$ trend. Based on this, \cite{Hen18} concluded that the optical and UV evolution in the 2016 eruption were similar to the previous ones, while the peculiarity of the 2016 eruption was reflected only in the X-rays. However, Figures~\ref{uv} and \ref{fig:lc_model} show that the evolution of $uvw2$ flux in 2016 differs from all other (subsequent) eruptions. The 2016 $uvw2$ light curve is fainter, similar to what was also noted for its soft X-ray counterpart (\S \ref{subsection:SSS phase}).

\subsection{Decreasing accretion rate}
In \S~\ref{sec:recurrence_nature}, we found that the recurrence period shows an increasing trend with time. A modest increase in the recurrence period would imply that either the accretion rate or the WD mass is decreasing over the years \citep{Hill16, wang18}. Light curve models provided by \cite{kato15} suggested the WD to be as massive as $\rm 1.38 ~ M_{\odot}$, accreting at a rate of $\rm 1.3~\times~10^{-7}~M_{\odot}~yr^{-1}$. Whereas \cite{dar17b} modelled the quiescent phase accretion disk using HST data and showed that the rate of mass accretion could be even higher at $\rm (0.6-1.4)~\times~10^{-6}~M_{\odot}~yr^{-1}$ considering a 50\% efficiency. On the other hand, the ejecta masses during each nova cycle (see \S~\ref{subsection:SSS phase} and \S~\ref{sec:spectra_model}) are $\rm \sim 10^{-8} M_{\odot}$. Thus, the net mass lost during each eruption is always less than the total mass gained between each eruption. As a result, the WD is growing in mass with time. The absence of Ne lines in the spectra indicates it to be a CO WD. Such WDs can reach $\rm >1.36~M_{\odot}$ only by accreting material. These inferences rule out the possibility of an increasing recurrence period due to decreasing WD mass. We suspect that the accretion rate has been slowly declining over the years (see Figure~\ref{fig:erupt_predict}), lengthening the time taken to reach the critical conditions required for thermonuclear runaway reactions to be initiated on the surface of the WD. 
The following could cause a gradual decrease in the accretion rate:
\begin{enumerate}
    \item The presence of starspots and increased activity in the secondary \citep{Hen18}. 
    \item The companion slowly running out of gas by supplying material to power the ``H flashes" for millions of years \citep{dar19}. 
    \item Orbital dynamics can also change accretion rates, especially in violent systems like M31N~2008-12a, where nova eruptions are frequent.
    \item Extent of the destruction of the accretion disk during each nova eruption decides the time taken to reform the accretion disk and resumption of accretion. Delayed accretion can also lead to a slowing down of the recurrence period. 
    \item A third body orbiting the M31N~2008-12a CV could perturb the binary motion, changing the accretion rate. Triple systems are known to produce exotic binaries; one such example is T Pyx \citep{kni22}.  
\end{enumerate}


\section{Summary}
This paper presents the evolution of $2017-2022$ eruptions of M31N~2008-12a in different wavelengths. The main results are summarised as follows.
\begin{enumerate}
    \item The linear decline post-maximum in the optical light curves is similar to that of the previous eruptions. The evolution of the UV light curve in the $2017-2022$ eruptions is also similar to the previous eruptions. A rapid decline since the maximum is followed by a plateau phase coincident with the SSS turn-on time. It then follows a secular decline with undulations before dimming beyond the detection limit. A UV rebrightening is also seen towards the end of the SSS phase.
    \item The SSS phase features are consistent with previously reported values. The mean SSS turn-on time and turn-off time are 5.1~$\pm$~0.6~days and 16.9~$\pm$~1.0~days since the eruption, respectively. The SSS phase shows X-ray variability, the most prominent being the dip $\sim$~11~days after the eruption. 
    \item The UV and soft X-ray flux are found to be anti-correlated at the peak of the SSS phase, which has not been reported before. This implies that both originate at the surface of the WD and could arise during the reformation of the partially disrupted accretion disk.
    \item Balmer, He, and N lines dominate the optical spectra.  H$\alpha$ velocities decelerate from $\sim$~5000~km~s$^{-1}$ within 1~day of eruption to $\sim$~2000~km~s$^{-1}$ at around 4~days after eruption, consistent with phase II of shock remnant development.
    \item The ejecta mass derived from $\rm t_{on}$ and \texttt{Cloudy} modelling is of the order of $\rm 10^{-8}-10^{-7}~M_{\odot}$, which is consistent with previous estimates derived using different techniques. Compared to the accretion rates derived in this work and previous studies, the ejecta mass is lower than the average mass accreted in a year, implying the WD is potentially increasing its mass.
    \item He abundance in the ejecta was found to be high at $\rm He/He_{\odot} \sim 2.5-3.1$, as is the case for most RNe.
    \item H$\alpha$ line morphology indicates an ejecta with an equatorial ring, a slow bipolar conical component, and an extended fast bipolar component along the line-of-sight resembling a jet-like structure. 
    Evidence of a cuspy feature in the light curves near the peak is seen as a general trend after the 2016 eruption in the $r'$ and $i'$ bands. Together with emission-line modelling, we conjecture the cusp is caused by jets present in the ejecta. The presence of jets in this system was suspected, and we provide strong evidence for its presence here. Such jets could be common in novae systems, especially RNe.
    \item We noticed that the recurrence period shows a weak tendency to increase with time, a sign of decreasing accretion rate.
    \item By comparing the recurrence period with binary evolution models, the mass of the WD is constrained to be $\rm > 1.36 ~ M_{\odot}$. However, we emphasize that none of the models could replicate the observed accretion rate determined in previous studies. Irrespective of that, a CO WD near the $\rm M_{Ch}$ and growing in mass is a good candidate for the single degenerate channel of Type Ia supernova explosions. 
    
\end{enumerate}

\section{acknowledgments}
We would like to thank the anonymous referees for their valuable comments that improved the quality of this work. 
We are grateful to the support staff and the observers at VBO and IAO. We take this opportunity to thank the TAC for the time allocation for ToO observations. We thank the staff of IAO, Hanle and CREST, Hosakote, that made these observations possible. The facilities at IAO, CREST, and VBO are operated by the Indian Institute of Astrophysics, Bangalore. 
We acknowledge the use of GROWTH-India telescope data. The GROWTH-India telescope (GIT) is a 70 cm telescope with a 0.7$^{\circ}$ field of view set up by the Indian Institute of Astrophysics (IIA) and the Indian Institute of Technology Bombay (IITB) with funding from the Indo-US Science and Technology Forum and the Science and Engineering Research Board, Department of Science and Technology, Government of India. It is located at the Indian Astronomical Observatory (IAO, Hanle). We acknowledge funding by the IITB alumni batch of 1994, which partially supports the operation of the telescope. Telescope technical details are available at \href{https://sites.google.com/view/growthindia/}{https://sites.google.com/view/growthindia/}.  
This work uses the SXT and UVIT data from the AstroSat mission of the Indian Space Research Organisation (ISRO). We thank the AstroSat TAC for allowing us ToO time to observe this nova from 2019-2022. We thank the SXT and UVIT payload operation centres for verifying and releasing the data via the ISSDC data archive and providing the necessary software tools. 
We acknowledge the use of public data from the {\it{Swift}} data archive. 
This work has also used software and/or web tools obtained from NASA's High Energy Astrophysics Science Archive Research Center (HEASARC), a service of the Goddard Space Flight Center and the Smithsonian Astrophysical Observatory. Kulinder Pal Singh (KPS) and G.C. Anupama (GCA) thank the Indian National Science Academy for support under the INSA Senior Scientist Programme.

\facilities{HCT:2m, JCBT:1.3m, GIT:0.7m, \textit{Swift} (UVOT and XRT), \textit{AstroSat} (UVIT and SXT)}

\software{\texttt{CCDLAB} \citep{postma17}, 
            \texttt{IRAF v2.16.1} \citep{tod93}, 
            \texttt{HEASOFT v6.25}, \texttt{XIMAGE v4.5.1}, and \texttt{XSELECT v2.4e} \citep{heasoft14}, 
            \texttt{XSPEC v12.12.0} \citep{arnaud96}, 
            \texttt{Python v3.6.6} \citep{python09},  
            \texttt{NumPy} \citep{NumPy20}, 
            \texttt{SciPy} \citep{SciPy20}, 
            \texttt{Pandas} \citep{mckinney10}, 
            \texttt{Matplotlib} \citep{Hunter07}, 
            \texttt{scikit-learn} \citep{sklearn11}, 
            \texttt{astroquery} \citep{gins19}
            }



\appendix

\bibliography{main}
\bibliographystyle{aasjournal}



\end{document}